\documentclass[aps,prb,reprint,superscriptaddress]{revtex4-1}
\pdfoutput=1

\usepackage{amsmath} 
\usepackage{graphicx} 
\usepackage{float}
\usepackage{color}

\newcommand{\figurewidth}{\columnwidth}

\renewcommand{\vec}{\bf}

\begin{document}

\title{Machine-learning based interatomic potential for amorphous carbon}
\author{Volker L. Deringer}
\email{vld24@cam.ac.uk}
\affiliation{Engineering Laboratory, University of Cambridge,
             Trumpington Street, Cambridge CB2 1PZ, UK}
\affiliation{Department of Chemistry, University of Cambridge,
             Lensfield Road, Cambridge CB2 1EW, UK}
\author{G\'a{}bor Cs\'a{}nyi}
\affiliation{Engineering Laboratory, University of Cambridge,
             Trumpington Street, Cambridge CB2 1PZ, UK}

\date{\today}

\begin{abstract}
  We introduce a Gaussian approximation potential (GAP) 
  for atomistic simulations of liquid and amorphous elemental carbon.
  Based on a machine-learning representation of the density-functional theory (DFT)
  potential-energy surface, such interatomic potentials
  enable materials simulations
  with close-to DFT accuracy but at much lower computational cost. 
  We first determine the maximum accuracy that any finite-range potential can
  achieve in carbon structures; then, using a novel hierarchical set
  of two-, three-, and many-body structural descriptors,
  we construct a GAP model that can indeed reach the target accuracy.  
  The potential yields accurate energetic and structural properties over
  a wide range of densities; it also correctly captures the structure of
  the liquid phases, at variance with state-of-the-art empirical potentials. 
  Exemplary applications of the GAP model to surfaces of  ``diamond-like'' tetrahedral
  amorphous carbon ({\em ta}-C)  are presented,
  including an estimate of the amorphous material's surface energy, and
  simulations of high-temperature surface reconstructions (``graphitization'').
  The new interatomic potential appears to be promising for realistic and accurate 
  simulations of nanoscale amorphous carbon structures.
\end{abstract}

 \maketitle

\section{Introduction}

\begin{figure*}
\centering
\includegraphics[width=17.8cm]{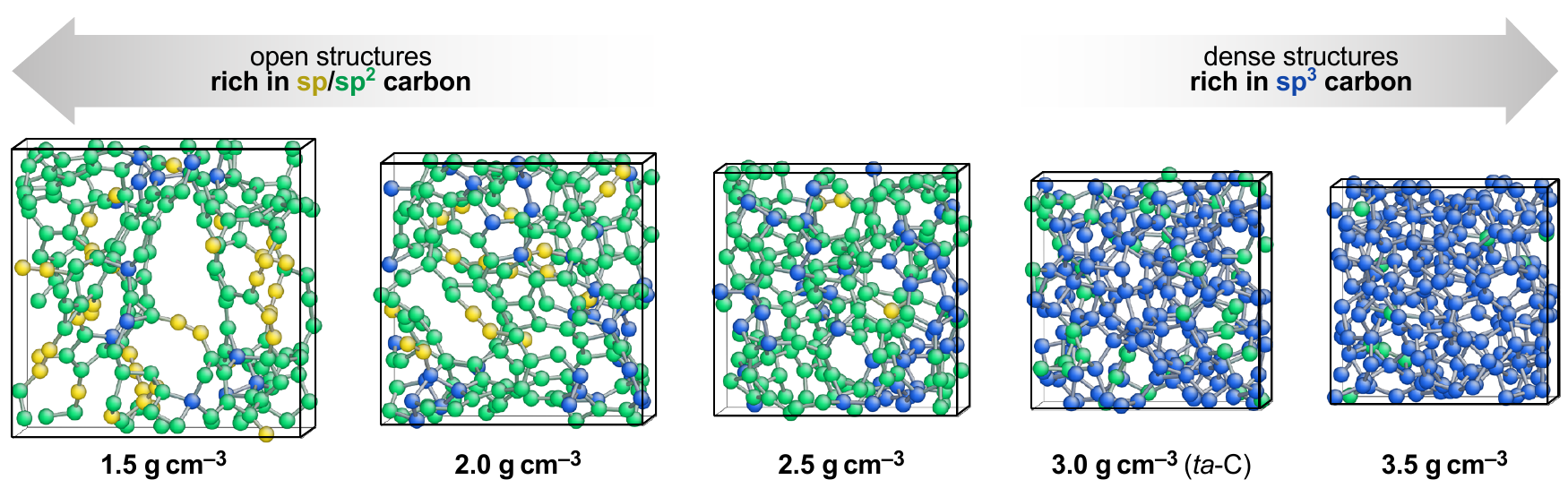}
\caption{\label{fig:structures}
         Exemplary $a$-C structures at various densities, obtained in 
         216-atom cells from DFT melt--quench simulations. Note the gradual
         transition from open to dense networks, and the coexistence of twofold
         (``sp''; yellow), threefold (``sp$^{2}$''; green), and fourfold (``sp$^{3}$'';
         blue) coordinated carbon atoms. 
         The open, low-density structures are metastable and on much further
         annealing will form more sp$^{2}$-rich networks; \cite{Powles2009, deTomas2016}
         here, on purpose, we focus on the as-quenched structures shown, 
         to assess as diverse local environments as possible.
         Bonds are drawn up to a 
         maximum interatomic distance of 1.85 \AA{}, 
         and coordination numbers are determined using the same cutoff.
         Structures were visualized using {\sc AtomEye}. \cite{AtomEye}
         }
\end{figure*}

Carbon is among the most intriguing elements due to its structural diversity,
and its solid-state forms range from diamond and graphite
via many more complex allotropes \cite{Hirsch2010, Georgakilas2015,
Hoffmann2016}
onward to amorphous phases ($a$-C). The atomic structures of $a$-C samples
depend strongly on density and are characterized by the coexistence
of threefold (``sp$^{2}$'') and fourfold bonded (``sp$^{3}$'')
carbon atoms. In this sense, low- and high-density forms of $a$-C are loosely
reminiscent of graphite and diamond, respectively, but the actual situation
is much more complex (Fig.\ \ref{fig:structures}).
``Tetrahedral amorphous'' carbon ($ta$-C), the dense, sp$^{3}$-rich form,
is of particular technological interest due to its attractive mechanical
properties. \cite{McKenzie1991, McKenzie1999, Robertson2002}

Atomistic simulations have long been providing useful insight into $a$-C materials.
\footnote{For an overview, see: N.~A.~Marks, in {\em Computer-Based Modeling
of Novel Carbon Systems and Their Properties: Beyond Nanotubes}, eds.
L.~Colombo, A.~Fasolino (Springer, Dordrecht, 2010), pp. 129--169.}
Many empirical interatomic potentials exist for carbon, from the original
Tersoff \cite{Tersoff1988} and Brenner \cite{Brenner1990}
formulations to more recent developments, including an environment-dependent
interaction potential (EDIP), \cite{Marks2000} improved reactive bond-order
(REBO) potentials, \cite{Brenner2002, Pastewka2008} 
or a recently re-parametrized reactive
force field (ReaxFF); \cite{Srinivasan2015}
a comprehensive comparative study of such potentials was very recently
carried out. \cite{deTomas2016}
These fast potentials make large-scale molecular-dynamics
(MD) simulations possible, and have been applied to
engineering problems such as fracture \cite{Jensen2015} 
or friction and wear of $ta$-C coatings;
\cite{Pastewka2010, *Pastewka2011} they are efficient enough to
perform thin-film deposition simulations, 
\cite{[{}][{, and references therein.}]Marks2002a} 
thus directly mirroring the atomic-scale processes in experiments.
Nonetheless, these potentials remain empirical in nature,
and may have serious shortcomings:
prominent examples are an underestimated
concentration of sp$^{3}$-bonded atoms in $ta$-C, \cite{Pastewka2008} 
and poor description of surfaces. A general problem of empirical potentials
is the inevitable compromise in accuracy for predicting different
material properties.    

On the other hand, seminal studies based on tight-binding schemes
\cite{Wang1993, *Wang1994, Frauenheim1993, Kohler1995} as well as
density-functional theory (DFT)
\cite{Galli1989, Drabold1994, Marks1996, McCulloch2000} 
early on afforded atomistic structure models of $a$-C,
and more recent DFT-MD studies dealt with applications in
photovoltaics \cite{Risplendi2014} or coatings. \cite{Music2016}
Furthermore, liquid carbon has been of interest---for example,
in first-principles studies of the diamond melting line which is difficult
to evaluate experimentally. \cite{Wang2005}
Despite their usefulness, however, DFT-based methods are limited to quite
small system sizes, and even with the computational power available nowadays,
they are limited in practice to a few hundred atoms. 
This makes many of the above scenarios simply inaccessible
to predictive DFT-quality simulations. 

To bridge the long-standing gap between these two realms, a novel class of
simulation methods has recently emerged
which is based on machine learning (ML).
The key idea is to map a set of atomic environments
{\em directly} onto numerical values for energies and forces; these quantities 
are ``trained'' from a large and accurate quantum-mechanical reference database but subsequently
interpolated using the ML algorithm. If training is successful, this makes 
atomistic simulations close to  quantum-mechanical accuracy accessible but requires
less computational effort by many orders of magnitude.
Recent implementations use high-dimensional artificial neural networks,
\cite{Lorenz2004, Behler2007, Artrith2016}
compressed sensing, \cite{Seko2015}
or Gaussian process regression. \cite{Bartok2010} 
Interatomic ML-based potentials have been developed for several
prototypical solids
\cite{Lorenz2004, Behler2007, Bartok2010, Eshet2010, Khaliullin2010, Artrith2011, 
Artrith2012, Szlachta2014, Seko2015, Artrith2016}
and applied, {\em e.g.}, in studies of phase transitions.
\cite{Behler2008, *Khaliullin2011}
We mention in passing that ML schemes are currently being developed to
estimate other fundamental properties of molecules and solids,
including atomization energies, \cite{Rupp2012, *Faber2016}
multipolar polarization, \cite{Darley2008, *Handley2009}
band gaps, \cite{Lee2016}
or NMR parameters. \cite{Rupp2015a, *Cuny2016}
A recent tutorial review of the field is in Ref.\ \citenum{Rupp2015}.

Previous ML potentials have been created for the crystalline
carbon allotropes diamond and graphite, \cite{Bartok2010, Khaliullin2010}
but as those were trained on a small region of configuration space,
they are not suitable for simulating  $a$-C. 
Indeed, the only reported ML potential dealing with amorphous
matter is a neural-network potential for the phase-change 
data-storage material GeTe
\cite{Sosso2012} that enabled large-scale simulations
of thermal transport \cite{Sosso2012a} and atomistic
processes during crystallization. \cite{Sosso2013}
Amorphous materials are structurally much more diverse than their
crystalline counterparts, and despite the lack of long-range translational
symmetry, their properties depend crucially on structural order on the local 
and intermediate length scales.\cite{Elliott1991} 
The required large unit cells and the long
relaxation times make it very difficult to use DFT simulations for 
amorphous materials of practical interest.
\cite{Sosso2012, Akola2012, *MRSBulletin}
The latter are hence particularly promising targets for high-quality ML
potentials. 

In this work, we introduce an interatomic Gaussian approximation 
potential (GAP) for condensed-phase elemental carbon, with 
particular focus on liquid and amorphous phases of various densities.
First, we systematically determine the maximum accuracy that any finite-range 
interatomic potential for carbon can achieve as a function of its neighbor
cutoff, independently of how it is fitted. 
Then, we show that our GAP does indeed
reach this accuracy, and furthermore provides reliable structural 
and topological data that agree well with the computationally
much more demanding DFT benchmarks.
Finally, we show predictions for energies and structures of $ta$-C surfaces,
which play a key role in fracture and wear processes.

\section{Theory}\label{sec:theory}

The Gaussian approximation potential (GAP) \cite{Bartok2010,Bartok2015} is an ML
approach to atomistic materials modeling, whereby an interatomic potential for the 
given material is
``trained'' from a database of reference quantum-mechanical data,
and is then used to interpolate energies and forces for 
arbitrary structures.
In order to make simulation of large systems feasible, the total energy is broken down
into a sum of local contributions, given by an local energy function $\varepsilon$. This function is  expanded in a basis set adapted to the input database;
it is generated using a kernel function, or similarity measure of neighbor environments.
The choice of this kernel (and the symmetries it obeys)
is critical for the success of any ML potential. \cite{Bartok2013}

Previous ML potentials for solids used a decomposition into atomic energies, and 
employed many-body descriptors to represent the atomic neighbor environment---comprising
all neighbors of an atom up to a given cutoff radius.
\cite{Bartok2010,Behler2011,Li2015,Botu2015,Shapeev2016}
However, for a complete description of these atomic environments one must fit the atomic
energy function in a high-dimensional space. This leads to poor ``extrapolation'',
that is, to a poor fit in regions of configuration space 
far away from any data points. 
A long simulation will likely find such regions---especially
at high temperatures, and/or when disorder is large. Indeed, in the present
case of $a$-C, we encountered problems early during training when using a
single many-body descriptor only: 
MD runs driven by such GAP models showed atoms aggregating at 
unreasonably small (sub-\AA{}) distances. This is a very general challenge
during the development of high-dimensional ML potentials, which carry the risk of erroneous
extrapolation behavior unless carefully tested and used.

In this work, we generalize the many-body GAP approach for solids:
we retain the many-body terms but augment
them with two- and three-body ``descriptors''---distances between 
atoms and angles in triplets. The latter terms hence
represent two- and three-body interactions
as in traditional (empirical) interatomic potentials, 
but now all descriptors and associated local-energy contributions are part of the
same ML framework.
Our starting point is thus the following expression for the total energy:
\begin{align}
 \label{eq:E-2b-3b-MB}
 \nonumber
 E =& \left( \delta^{\rm (2b)} \right)^{2} \sum_{i \, \in \, {\rm pairs}}
        \varepsilon^{\rm (2b)} ({\vec{q}}^{\rm (2b)}_{i}) \\
     +& \left( \delta^{\rm (3b)} \right)^{2} \sum_{j \, \in \, {\rm triplets}}
        \varepsilon^{\rm (3b)} ({\vec{q}}^{\rm (3b)}_{j}) \\
 \nonumber
     +& \left( \delta^{\rm (MB)} \right)^{2} \sum_{a \, \in \, {\rm atoms}}
        \varepsilon^{\rm (MB)} ({\vec{q}}^{\rm (MB)}_{a}) 
\end{align}
where ``2b'', ``3b'', and ``MB'' denote two-, three-, and many-body interactions,
respectively. This is similar in spirit to the recently introduced Moment Tensor 
Potentials,\cite{Shapeev2016} and also to another scheme that uses a parametric 
two-body term in combination with a neural network that describes the many-body 
interactions\cite{Dolgirev2016}. 

In the above expression, the $\delta$ are scaling parameters, and each corresponds to
the distribution of
energy contributions a given interaction term has to represent.
We choose the largest value for the 2b terms, which describe the largest share
of the total energy; on top of that, we add a 3b term, and finally
the many-body term with the smallest $\delta^{(d)}$.

The local energy corresponding to each descriptor 
$d \in \{ \textrm{2b},\textrm{3b},\textrm{MB} \}$ 
is given by a linear combination of kernel functions \cite{Bartok2010}
\begin{equation}
 \label{eq:epsilon}
 \varepsilon^{(d)}({\vec{q}}^{(d)}) = \sum_{t=1}^{N^{(d)}_{t}} \alpha^{(d)}_{t} K^{(d)}({\vec{q}}^{(d)}, {\vec{q}}^{(d)}_{t}),
\end{equation}
where $t$ denotes one of $N_{t}$ training configurations ${\vec{q}}_{t}$,
each of which attains a weighting coefficient $\alpha_{t}$ during fitting,
and $K$ is a covariance kernel which quantifies how similar the input
configuration ${\vec{q}}$ is to the  $t$-th training configuration, ${\vec{q}}_{t}$.
In practice, we sparsify the representation and only allow the sum to range over a
number of ``representative points'' drawn from the full
training database ($N_{t} \ll N_{\rm full}$). The number of representative points
differs for each descriptor and must be carefully controlled during training.

Both for 2b and 3b contributions, we use a squared exponential
kernel, \cite{Bartok2010}
\begin{equation}
 K^{(d)} \left({\vec{q}}^{(d)}_{i}, {\vec{q}}^{(d)}_{t} \right) =
  \exp \left[ -\frac{1}{2} \sum_{\xi} 
  \frac{(q_{\xi, i}^{(d)} - q_{\xi, t}^{(d)})^{2}}{\theta_{\xi}^{2}} \right],
\end{equation}
where $\xi$ is an index running over the components of the descriptor vector
${\vec{q}}^{(d)}$.
In the case of pairs, the descriptor has one single scalar component (namely, the
distance $r_{12}$ between the two atoms involved):
\begin{equation}
{q}^{\rm (2b)} = \left|{\vec{r}}_2-{\vec{r}}_1 \right| \equiv r_{12} \; ;
\end{equation}
for triplets, we do not
directly use the natural coordinates $r_{12}$, $r_{13}$, and $r_{23}$, but
a different form to enforce symmetry over permutation of 
the neighbor atoms 2 and 3:\cite{Bartok2015}
\begin{equation}
 {\vec{q}}^{\rm (3b)} =
 \begin{pmatrix} r_{12}+r_{13} \\ \left(r_{12}-r_{13}\right)^{2} \\ r_{23} \end{pmatrix}.
\end{equation}

Note that with this choice of descriptors, the first term in Eq.~\ref{eq:E-2b-3b-MB}
is equivalent to a pair potential, and the second is a generic three-body potential,
but in the GAP framework
both do not impose constraints on the specific functional form. 

For the many-body term, we use the recently introduced 
Smooth Overlap of Atomic Positions (SOAP) \cite{Bartok2013} descriptor,
which has proven successful in generating
GAP models for tungsten,\cite{Szlachta2014} 
in classifying diverse molecular and solid-state
structures,\cite{De2016} and very recently in constraining
structural refinements of amorphous Si.\footnote{M.~J.~Cliffe, A.~P.~Bart\'o{}k,
R.~N.~Kerber, C.~P.~Grey, G.~Cs\'a{}nyi, and A.~L.~Goodwin, arXiv:1609.00668 (2016).}
We briefly review the
most pertinent features; detailed formulae and derivations are in Ref. 
\citenum{Bartok2013}.
SOAP starts from the neighborhood density of a given atom $a$, defined as
\begin{equation}
  \rho_{a}({\vec{r}}) = \sum_{b} 
  \exp \left[ - \frac{ ({\vec{r}} - {\vec{r}}_{ab})^{2} }{ 2 \sigma_{\rm at}^{2}} 
       \right]\times f_\mathrm {cut}(r_{ab}),
\end{equation}
where the sum is over neighboring atoms, and the cutoff function $f_\mathrm{cut}$,
which ensures compact support, goes smoothly to zero at $r_{\rm cut}$ over a
characteristic width $r_{\Delta}$. 
The parameter $\sigma_\text{at}$ ultimately controls the smoothness of the potential.
The neighbor density is expanded into a local basis of
orthogonal radial basis functions $g_{n}$ and spherical harmonics $Y_{lm}$,
\begin{equation}
  \label{eq:soap_rho_expansion}
  \rho_{a}({\vec{r}}) = \sum_{nlm} c^{(a)}_{nlm} \, g_{n}(r) Y_{lm}(\hat{\vec{r}}),
\end{equation}
and the expansion coefficients are used to form  the spherical power spectrum, 
\begin{equation}
 p^{(a)}_{nn'l} =\sqrt{\frac{8 \pi^{2}}{2l+1}} \sum_{m} \left( c^{(a)}_{nlm} \right)^{\ast} c^{(a)}_{n'lm},
\end{equation}
which is invariant both to permutations over neighbors and to 3D rotations of the
neighbor environment. We use the elements of a finite truncation of the power spectrum
(up to $n \leq n_\text{max}$ and $l \leq l_\text{max}$) as components of the many-body 
descriptor  vector ${\vec{q}}_a^\mathrm{(MB)}$, which furthermore
is normalized to have unit length.

The kernel function for the SOAP term is the simple dot product,
\begin{equation}
 k({\vec{q}}^{\rm (MB)}_{a}, {\vec{q}}^{\rm (MB)}_{t}) = 
 \sum_{nn'l} p^{(a)}_{nn'l} \, p^{(t)}_{nn'l} = 
 {\vec{q}}^{\rm (MB)}_{a} \cdot {\vec{q}}^{\rm (MB)}_{t},
\end{equation}
and we find it advantageous to raise it to a small integer power
for a sharper distinction between different environments. This gives the final kernel
\begin{equation}
 K^{\rm (MB)}({\vec{q}}^{\rm (MB)}_{a}, {\vec{q}}^{\rm (MB)}_{t}) = 
 \left| {\vec{q}}^{\rm (MB)}_{a} \cdot {\vec{q}}^{\rm (MB)}_{t} \right|^{\zeta}.
\end{equation}

This dot product kernel is a natural choice to use with 
the power spectrum descriptor, as it makes the kernel equivalent
(up to normalization) to the integrated overlap of the original neighbor densities, 
\begin{equation}
  \label{eq:soap-k}
   \int d\hat{R}  \left|  \int \rho_{a}({\vec{r}})
  \rho_{t}(\hat{R} {\vec{r}}) \right|^{2}.
\end{equation}

The expression for the total energy in our GAP model is therefore given by
\begin{align}
  \label{eq:gap-total-energy}
  \nonumber E =& 
  \left( \delta^{\rm (2b)} \right)^{2} \sum_{i} \sum_{t} \alpha^{\rm
(2b)}_{t}
            K^{\rm (2b)} \left({\vec{q}}^{\rm (2b)}_{i}, {\vec{q}}^{\rm (2b)}_{t}
\right) \\
  +& \left( \delta^{\rm (3b)} \right)^{2} \sum_{j} \sum_{t} \alpha^{\rm
(3b)}_{t}
            K^{\rm (3b)} \left({\vec{q}}^{\rm (3b)}_{j}, {\vec{q}}^{\rm (3b)}_{t}
\right) \\
  \nonumber
  +& \left( \delta^{\rm (MB)} \right)^{2} \sum_{a } \sum_{t} \alpha^{\rm
(MB)}_{t}
            K^{\rm (MB)} \left({\vec{q}}^{\rm (MB)}_{a}, {\vec{q}}^{\rm (MB)}_{t}
\right), 
\end{align}
where all fitting coefficients $\alpha$ enter linearly, and therefore we can obtain
them simply using linear algebra. This is in contrast with the difficult nonlinear
parameter optimization required both for traditional interatomic potentials and
for some other ML schemes, e.g., artificial neural networks. 

The above discussion does not include the prescription for obtaining the linear
fitting coefficients. In practice, this is complicated due to the fact that the
quantum mechanical data is only available in the form of {\em total} energies, atomic
forces, and virial stresses. The full formalism simultaneously includes
sparsification, multiple energy terms, and fitting to total energies and their
derivatives; is is given elsewhere.\cite{Bartok2015}

\begin{figure}[tb]
\centering
\includegraphics[width=\figurewidth]{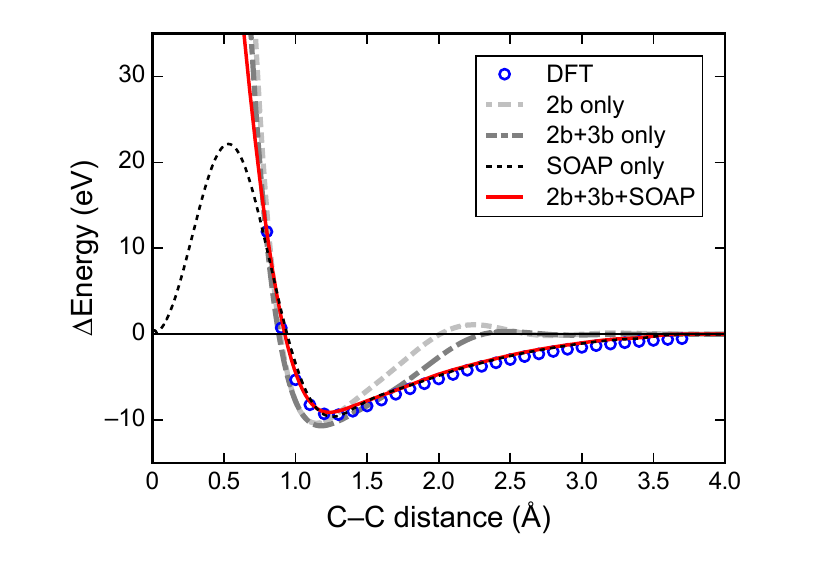}
\caption{\label{fig:dimer_potentials}
Potential-energy scans for an isolated carbon dimer. This plot, with 
DFT data as reference (blue), allows us to assess the use of
different structural descriptors: all three combined are needed for
a high-quality fit (see text).}
\end{figure}

To illustrate the role of the combined descriptors, we use different 
(and increasingly complex) GAP models to compute the potential-energy
curve for an isolated carbon dimer; these models have been fitted to 
the full bulk and surface training set described below that additionally
incorporates DFT data points between 0.8 and 3.7 \AA{} in small increments.
The results are summarized in Fig.\ \ref{fig:dimer_potentials}:
GAP models using 2b descriptors only, or a combination of 2b+3b, reproduce
the minimum and the repulsion at small C--C distances reasonably well, but
the longer-range behavior is not yet correctly described. An interesting
result is seen when using a many-body descriptor {\em only}: the fit is
very good for the region where data points are provided (blue circles), but
shows unphysical behavior at $r < 0.8$ \AA{}; this can, and will, then lead to
bad extrapolation in practical simulations. By contrast, a GAP model
combining all three descriptors (Eq.\ \ref{eq:gap-total-energy}) gives a highly
satisfactory result (red line in Fig.\ \ref{fig:dimer_potentials}). 

\section{Computational methods}

\subsection{General protocol for melt--quench simulations}

Structural data were obtained from melt--quench MD,
following protocols that are well established for $a$-C.
\cite{Galli1989, Marks1996}
Initial simulations were performed in the DFT framework,
subsequent ones with GAP, but both employed the same temperature
protocol. For each simulation, an (unstable) simple-cubic lattice of carbon
atoms was generated at the appropriate density and held at a constant temperature of 
9000 K for 3 ps. The simulation
cell was then held in the liquid state at 5000 K (3 ps),
quenched with an exponentially decaying temperature profile (0.5 ps),
and finally annealed at 300 K (3 ps).
The timestep was 1 fs in all MD simulations.

\subsection{DFT-based ({\em ``ab initio''}) molecular dynamics}

Structures for initial training, as well as benchmarks for
{\em a}-C properties, were generated using DFT-based {\em ab initio}
MD, using the {\sc Quickstep} scheme 
and a stochastic Langevin thermo\-stat \cite{Bussi2007} 
as implemented in {\sc cp2k}.
\cite{[{}][{; J. Hutter, M. Iannuzzi, F. Schiffmann, and J. VandeVondele,
Wiley Interdiscip.\ Rev.--Comput.\ Mol.\ Sci. {\bf 4}, 15 (2014); 
www.cp2k.org.}]VandeVondele2005, Kuehne2007}
Electronic wavefunctions were described at the $\Gamma$ point
using a mixed-basis scheme with
Goedecker--Teter--Hutter pseudopotentials \cite{Goedecker1996}
and a cutoff energy of 250 Ry. Double-$\zeta$ quality basis functions 
were used for the carbon 2s and 2p levels. 

Exchange and correlation were treated in the local density
approximation (LDA), \cite{DFT-KS} both during {\em ab initio} MD and
training-data generation. This functional, despite its simplicity,
has long been used for atomistic simulations of $a$-C and
is still the {\em de facto} standard for many current applications. 
\cite{Pastewka2008, Risplendi2014, Music2016}
Further work may be concerned
with the application of higher-level DFT methods, such as computationally
much more expensive hybrid
functionals, or the implementation of dispersion corrections---these
will likely be interesting additions to the GAP framework, but are 
beyond the scope of the present study.

\subsection{Construction of the training database}

Our training database contains structural snapshots from {\em ab initio} MD
and also, as it is iteratively extended, from GAP-driven simulations. No matter
how generated, all structures are then subjected to single-point
DFT-LDA computations, to yield well-converged energies and forces
for training. 
This was done using CASTEP,\cite{CASTEP}
with dense reciprocal-space meshes (maximum spacing 0.03 \AA{}$^{-1}$),
\cite{MonkhorstPack} a 650 eV cutoff for plane-wave expansions,
and an extrapolation scheme to counteract finite-basis errors. \cite{FBSC}
Gaussian smearing of 0.1 eV width was applied to
electronic levels. The halting criterion for SCF iterations was
$\Delta E < 10^{-8}$ eV.

Initial training data were computed for snapshots from {\em ab initio} MD
melt--quench trajectories, and a preliminary GAP was fitted to those data.
The resulting potential reproduced the structure of the
9000 K liquid well, that of the 5000 K liquid satisfactorily, but not yet 
that of the amorphous phase. In retrospect, this is easily
understood: the 9000 K liquid is highly diffusive, and so one single
3~ps trajectory apparently contains sufficiently different atomic environments
to sample configuration space during training. A quenched amorphous structure, by 
contrast, is essentially one single snapshot with thermal fluctuations
but no major changes in
connectivity. Training from DFT data alone would thus incur significant 
expense, as each uncorrelated $a$-C sample would require a {\em full} melt--quench 
trajectory (9500 steps) of which only the last snapshot were of use.  

Instead, an initial GAP was used to generate liquid structures at 5000 K,
which were then briefly 
re-equilibrated (500 steps) and quenched (500 steps) using {\em ab initio} MD.
This was done for ten uncorrelated structures each at 2.0, 3.0, 3.25,
and 3.5 g cm$^{-3}$, thus placing more emphasis  on high-density amorphous 
phases which are richer in
tetrahedral (``sp$^{3}$'') motifs and thus structurally
most different from the liquids.

The resulting, amended database was used to train a new GAP, which was
further extended iteratively  by performing melt--quench
simulations {\em fully} driven by the previous GAP version, as is common
practice in the development of ML potentials. \cite{Sosso2012, Szlachta2014}
Thereby, all GAP-MD simulations were carried out using a Langevin thermostat
as implemented in {\tt quippy} (www.libatoms.org), and 
the same temperature profiles as in the {\tt cp2k} simulations.
A typical protocol included the generation of 100 independent 
structures at densities of 1.5--3.5 cm$^{-3}$,
with system sizes of 27--125 atoms.
For one or more snapshots from each trajectory,
a single-point DFT computation was performed and the results were
included in the next round of training. 

To add amorphous surfaces to the training set,
we generated $ta$-C structures using GAP, and from these
created slabs by adding vacuum regions.
In parallel {\em ab initio} MD runs,
amorphous slabs were briefly heated at up to 5000 K, and structures from both procedures 
were added to the database. We reiterate that it is not problematic
to generate the training {\em structures} with different techniques,
\footnote{This can even turn out as an advantage---namely, if two techniques
lead to different and simultaneously relevant regions of configuration space.}
as their {\em energies and forces} are re-computed
using the same reference method (tightly converged plane-wave DFT).
  
Once the training database of liquid and amorphous structures had been generated,
it was further extended by including randomly distorted unit
cells of the crystalline allotropes, diamond and graphite. 
This combined database was then split into a training and a test set (90:10);
the latter was not included in the fit but used for validation. 
Finally, DFT data for an isolated dimer were added 
(cf.\ Fig.\ \ref{fig:dimer_potentials}).
A full description of the training database is provided as
Supplementary Information.

\begin{table}[tb]
\centering
\caption{Key parameters for the  GAP model created in this work
(see Sec.\ \ref{sec:theory} for definitions).}
\label{tab:GAP-params}
\begin{tabular}{lccc}
\hline
\hline
                       & $\;$ 2-body $\;$ & $\;$ 3-body $\;$ & $\;$ SOAP $\;$ \\
\hline
$\delta$ (eV)$^{a}$    & 5.0      & 0.3 & 0.1  \\
\hline
$r_{\rm cut}$ (\AA{})  & 3.7      & 3.0   & 3.7  \\
$r_{\Delta}$  (\AA{})  &          &       & 0.5  \\
\hline
$\sigma_{\rm at}$ (\AA{}) &       &       & 0.5  \\
$n_{\max}$, $l_{\max}$ &          &       & 8    \\
$\zeta$                &          &       & 4    \\
\hline
Sparsification         & Uniform  & Uniform & CUR \\
\hline
$N_{t}$ ($a$-C bulk)     & & 125   & 2500 \\
$N_{t}$ ($a$-C surfaces) & &  50   & 1000 \\
$N_{t}$ (crystalline)    & &  25   &  500 \\
\hline
$N_{t}$ (dimer)          & &       &   30 \\
\hline
$N_{t}$ (total)   & 15       & 200 & 4030 \\
\hline
\hline
\multicolumn{4}{p{7.5cm}}{\footnotesize
$^{a}$For the 2b and 3b descriptors, when specifying training input,
the $\delta$ given here is divided by the expected number of
pairs or triplets an atom is involved in.}
\end{tabular}
\end{table}

\subsection{GAP model fitting}

Values for the above GAP parameters as used in the present work are given in
Table~\ref{tab:GAP-params}. 
Furthermore, the regularization parameters of the Gaussian process corresponding
to the expected errors were as
follows. For liquid and amorphous structures
we set 0.002 eV (energies) and 0.2 eV \AA{}$^{-1}$ (forces); for the crystalline
forms, we multiplied both values by 0.1, and additionally included virials in the
training, with an expected error of 0.2 eV.
Sparsification was done with the CUR method\cite{Mahoney2009} for the SOAP kernel,
whereas a simple uniform grid of basis function locations was used for the 2b and 3b terms.
In the following, unless specified otherwise, ``GAP model'' refers to one
with all three terms (2b, 3b, and SOAP).
The potential files are freely available at http://www.libatoms.org.

\section{Results and discussion}

\subsection{Locality and target accuracy}\label{sec:locality}

A central assumption of all interatomic potentials is that of {\em locality}:
the energy associated with a given atom or bond depends on its immediate
environment ($\varepsilon_{i} \equiv \varepsilon({\vec{q}}_{i})$),
but not on atoms further away than a given cutoff radius
(ignoring electrostatic terms and van der Waals corrections for the moment).
A similar assumption follows directly for the forces on atoms.
While this approximation is often made implicitly, and tacitly, in the
development of empirical potentials, their ML-based counterparts aim at
{\em quantitative} energy and force accuracy with respect to the reference
potential-energy surface, and so at the outset we must numerically determine
how well the above assumption holds.
This question is very general, and likely relevant beyond the present study.

\begin{figure}[tb]
\centering
\includegraphics[width=\figurewidth]{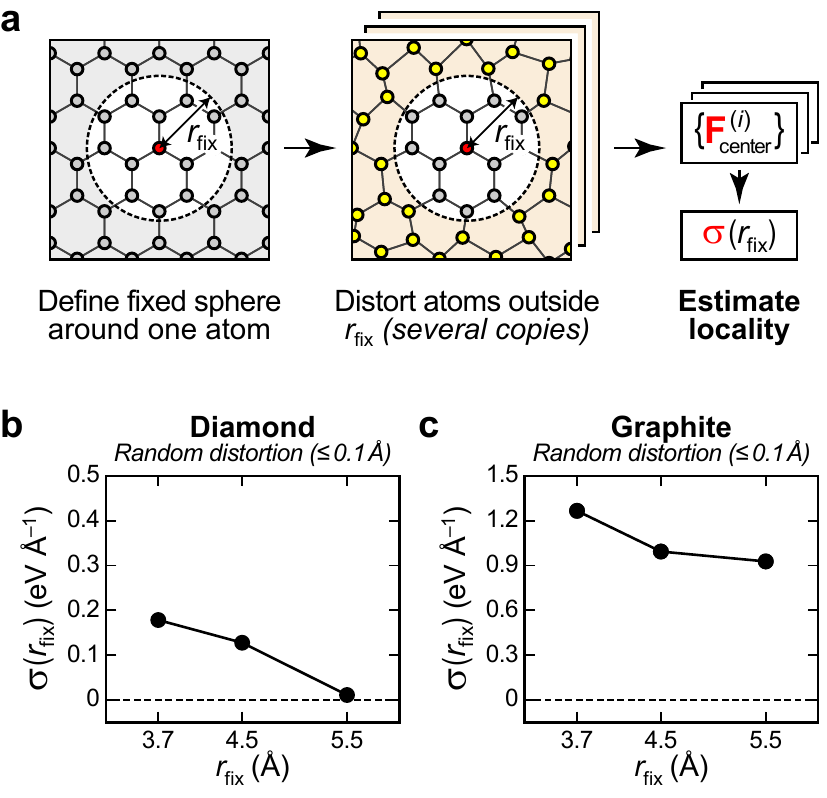}
\caption{\label{fig:locality_combined_cryst}
         (a) Schematic overview of the procedure used here for locality tests.
         (b,c) Results for diamond and graphite, respectively, obtained by
         displacing all atoms outside $r_{\rm fix}$ randomly by up to 0.1 \AA{}
         and inspecting the standard deviation of the force on the central
         atom.}
\end{figure}  

Quantum-mechanical models such as DFT are inherently nonlocal: they do not allow
for a unique partitioning of the total energy into a sum of local terms. Nonetheless,
quantum models of electronic structure are {\em\ nearsighted},\cite{Kohn1996}
which means that the reduced one-particle density matrix decays strongly
(at least under the assumption of screening, for insulators, and in general at 
finite temperature).\footnote{F.~Q.~Nazar and C.~Ortner, arXiv:1509.06753 (2016).}
This implies locality in the atomic forces, which we quantify as the decay of 
the dependence of an atomic force on a neighboring  atom's position
as the distance between the two atoms grows. A direct manifestation of this
is the decay of the dynamical matrix or Hessian. 

Using our reference quantum-mechanical method, we can calculate the above decay
of the dependence of the atomic forces, and thus determine a bound on force locality.
Conversely, this gives a bound on the force accuracy of {\em\ any} interatomic
potential model based on a local energy decomposition. We stress again that all this 
applies only for materials without strong polar interactions---or for models from
which such polar interactions have been substantially subtracted.

\begin{figure}[tb]
\centering
\includegraphics[width=\figurewidth]{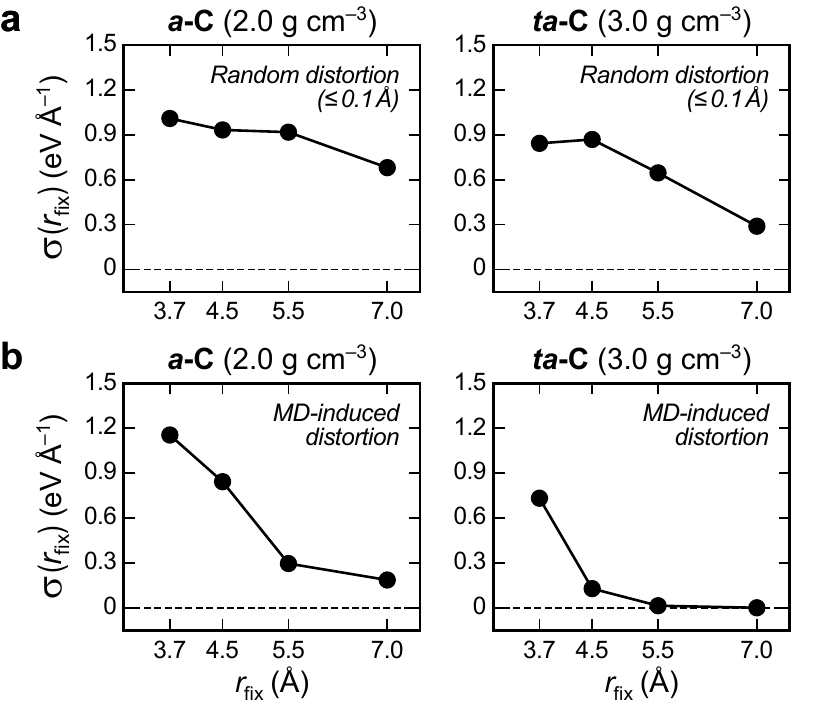}
\caption{\label{fig:locality_combined_amo}
         Locality tests for $a$-C structures.
         (a) Force locality in low- and high-density forms,
         evaluated in the presence of a small
         distortion that preserves the major topological features of the amorphous
         network. In both panels, data have been collected over three
         structural models; for each, ten atoms were randomly sampled as sphere
         centers, and five independent distortions were created for each central
         atom. 
         (b) Same but for a large distortion 
         induced by MD 
         at very high temperature such as to erase any structural 
         memory of the initial cell outside the fixed sphere.}
\end{figure}

The procedure, as previously employed by Bart\'o{}k {\em et al.}, \cite{Bartok2010}
is sketched in Fig.\ \ref{fig:locality_combined_cryst}a:
we select one atom in a given simulation cell and define a sphere of 
radius $r_{\rm fix}$ around this atom in that the atoms are fixed. 
We then create many structures which differ in the positions of the atoms
{\em\ outside} the fixed sphere, and for each calculate the force on the
central atom using 
DFT.\footnote{For the locality tests, we used CASTEP
as described in the Methods section; the reciprocal-space grids
were reduced to make the computational workload
tractable.} Locality is then characterized by plotting the standard
deviation of this force as a function of $r_\text{fix}$.  

We first consider the crystalline allotropes and begin by introducing rather modest
distortions, moving all atoms outside $r_{\rm fix}$ randomly by up
to $0.1$ \AA{}.
Diamond exhibits strong locality (Fig.\ \ref{fig:locality_combined_cryst}b):
the overall force deviations due to displacements outside the spheres are small,
and they gradually vanish and are practically zero at $r_{\rm fix} = 5.5$
\AA{}. Graphite, by stark contrast,
is highly non-local (Fig.\ \ref{fig:locality_combined_cryst}c):
the force deviations are much larger than in diamond,
and they do not decay as rapidly.

Turning now to the locality in amorphous carbon, we focus on two representative
densities: a low-density form (2.0 g cm$^{-3}$) and an approximant of
dense $ta$-C (3.0 g cm$^{-3}$), and again we start by randomly
displacing atoms by up to 0.1 \AA{} (Fig.\ \ref{fig:locality_combined_amo}a).
Qualitatively, the results seem to be in line with those for the crystalline
phases: the
more sp$^{2}$-rich form (2.0 g cm$^{-3}$) clearly shows lower
locality.\footnote{One is tempted to correlate this with the more
metallic nature of lower-density $a$-C (see, e.g.,
Ref.\ \citenum{Risplendi2014}); a definitive answer would require sampling
over a much larger number of structures, though, especially given the large
error bars involved.}
Due
to the coexistence of sp$^{2}$/sp$^{3}$ motifs in the amorphous forms,
however, there is no clear-cut distinction between the two system sizes,
and $ta$-C retains a notable degree of nonlocality. 

The displacements so far have perturbed the atoms outside
$r_{\rm fix}$,  but the models still
retain a ``memory'' of the initial structure even outside the fixed sphere. 
We therefore next perform Tersoff MD,\cite{Tersoff1988} starting with velocities
that correspond to a very high temperature, and let the system evolve for 1 ps,
again keeping the central sphere fixed. This leads to a more local picture
(Fig.\ \ref{fig:locality_combined_amo}b), especially for the dense
``diamond-like'' form; nonetheless, the overall $\sigma$(3.7 \AA{}) values
in the latter are much larger than in the crystalline form .

Summarizing, diamond shows the strong force locality expected for a 
covalent semiconductor; graphite, by contrast, is highly nonlocal.
The latter holds for the amorphous phases as well, more pronounced so at low
density. As a ballpark measure, for an $a$-C potential
with a cutoff radius of 3.7 \AA{}, we estimate the lowest achievable
standard deviation of force components to be $\approx 1$ eV \AA$^{-1}$ (Fig.\ \ref{fig:locality_combined_amo}).
One might increase $r_{\rm cut}$ up to 7.0 \AA{}, which
is expected to lower the standard deviation to
$\approx 0.7$ eV \AA$^{-1}$, but the tradeoff
in terms of much greater computational expense (both during training and
application of the GAP) does not seem to justify this.
Hence, all that follows will be done in the framework of modest $r_{\rm cut}$
values as given in Table \ref{tab:GAP-params}. 

\begin{figure}[bt]
\centering
\includegraphics[width=\figurewidth]{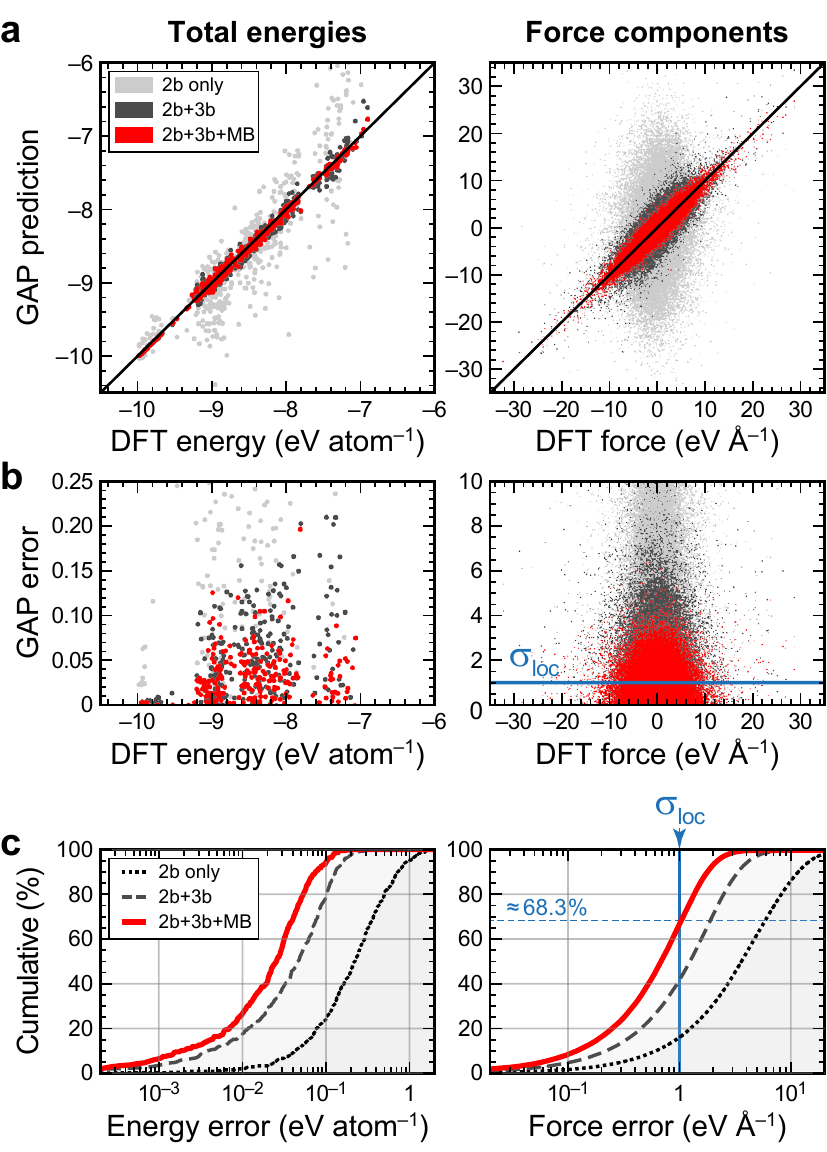}
\caption{\label{fig:scatter_das}
(a) Scatterplots of DFT-computed and GAP-predicted total energies
({\em left}; relative to a free singlet atom) and force components ({\em right})
on a test set of 450~configurations.
Results are shown for hierarchical GAP models 
with different combinations of descriptors (Sec.\ \ref{sec:theory}).
(b) Absolute errors of the respective quantities, similarly resolved according to
different sets of descriptors.
For the force components ({\em right}), akin to Ref.\ \citenum{Bartok2010},
an estimate of the maximum achievable standard deviation as judged from locality
tests (Sec.\ \ref{sec:locality}) is indicated by a blue line.
(c) Cumulative distributions: a given point ($x,y$) on the curve indicates 
that $y$ percent of all structures have an error equal to or below $x$. The
standard deviation estimated from locality tests, $\sigma_{\rm loc}$,
which should enclose $\approx 68.3\%$ of the GAP force component
errors, is indicated in blue:
indeed, the GAP model with combined 2b, 3b, and MB descriptors (red line)
does reach this accuracy.}
\end{figure}

\subsection{Energies and forces}

With the target errors for a finite-range potential established,
we can now analyze the quality of our GAP. We therefore test how much the
predicted energies and forces deviate from DFT reference values.
Again, we assess different combinations of structural descriptors,
and thereby illustrate how hierarchical
GAP models can achieve increasing accuracy. Correlation plots
of energies and force components already make this clear
(Fig.\ \ref{fig:scatter_das}a): using
the 2b descriptor only, there is a certain degree of correlation between
the DFT and GAP energies, but with much scatter,
and there is essentially no correlation between DFT and GAP force
components (light gray).
A 2b+3b model is clearly better (dark gray), but ultimately SOAP must be added (red) 
to achieve the accuracy limit imposed by nonlocality (Fig.\ \ref{fig:scatter_das}b).
Figure \ref{fig:scatter_das}c shows the errors as cumulative
distributions: the curves move left (toward
lower errors) and up (to a higher degree of confidence) as successively more
complex descriptors are added to the GAP model. 

\begin{figure}[tb]
\centering
\includegraphics[width=\figurewidth]{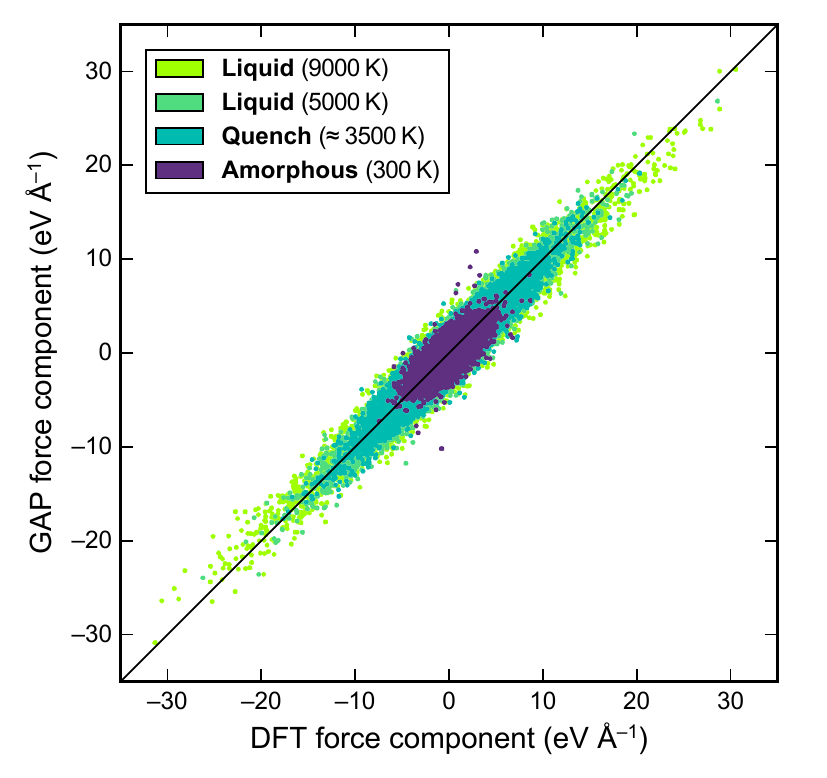}
\caption{\label{fig:stages}
 DFT-computed versus GAP-predicted force
 components in a set of 125-atom snapshots of liquid and amorphous
 carbon, emphasizing the overall magnitude of forces the GAP has
 to learn at various parts of the melt--quench trajectories. }
\end{figure}

For such a heterogenous training database, it is interesting to further 
break down the GAP's performance according to configuration types:
slightly distorted diamond configrations will 
be easier for an ML potential to fit than 
disordered liquid carbon. Indeed, looking back at Fig.\ \ref{fig:scatter_das}
shows that the training points with lowest overall energy show the lowest
fitting errors; these are precisely the crystalline structures.

In Fig.\ \ref{fig:stages}, we show the distribution of errors
for configurations coming from different stages of the melt--quench cycles.
We investigate a set of 100 uncorrelated $a$-C structures, with 125 atoms each and
randomized densities over the range 1.5--3.5 g cm$^{-3}$, created using 
GAP-MD and subsequently analyzed using DFT.
From each melt-quench trajectory, we take one configuration at each key
step---that is, one from the 9000 K and one from the 5000 K liquid, 
one during quenching, and the final one from the quenched amorphous sample.
The force errors  are very similar for
all parts of the trajectory, but the {\em absolute} magnitude of forces
is much different; hence, in {\em relative} terms, the GAP performs much better
for forces in the liquid than in the amorphous phases. 
A detailed numerical analysis is in Table \ref{tab:rms-errors}. We
estimate how widely the absolute DFT force components are distributed
 by giving their 95-th percentile value $P_{95}$. 
We then divide the GAP force component error by
$P_{95}$; the lower this ratio, the better.
For melt--quench simulations, the situation appears favorable:
as the structure is
``frozen in'' during quenching, the topology (say, the sp$^{3}$ count)
of the amorphous phase is largely determined by a correct description of
the liquid.

\begin{table}[tb]
\centering
\caption{Energy and force RMS errors of our GAP, 
computed for a set of 125-atom structures (cf.\ Fig.\ \ref{fig:stages}),
and also for the crystalline structures from the test set.
Percentile values $P_{95}$ for the absolute
DFT values are given; these measure the range of data the GAP has to ``learn''.}
\label{tab:rms-errors}
\begin{tabular}{lcccccc}
\hline
\hline
  & Energy & & \multicolumn{3}{c}{Force components} \\
  & (eV)   & & \multicolumn{3}{c}{(eV \AA{}$^{-1}$)} \\
\cline{2-6}
  & RMS & & RMS & $P_{95}$ &  \\
  & (GAP) & & (GAP) & (DFT) & Ratio \\
\hline
Liquid (9000 K) & 0.041 & & 1.27 & 6.52 & 0.19 \\
Liquid (5000 K) & 0.031 & & 1.12 & 5.68 & 0.20 \\
Quench          & 0.023 & & 1.07 & 5.06 & 0.21 \\
Amorphous       & 0.018 & & 0.94 & 2.23 & 0.42 \\
\hline
Crystalline     & 0.002 & & 0.10 & 1.32 & 0.08 \\
\hline
\hline
\end{tabular}
\end{table}

\subsection{Structural properties}

From energy and force evaluations, we now move on to probe physical
properties as predicted by our GAP.
 Table \ref{tab:diamond}
compares its performance to DFT reference data for the diamond structure.
Here and in the following, we will also make comparison
to a state-of-the-art empirical potential, namely, a screened variant
of the Tersoff potential developed by Pastewka and coworkers. \cite{Pastewka2008,
Pastewka2013} Similar potentials have been successfully applied in recent
studies both to graphene \cite{Klemenz2014} and to $ta$-C, \cite{Kunze2014}
and are   faster than  GAP by about a factor of 50.

The lattice parameter $a$ of diamond is accurately reproduced by the GAP;
the bulk modulus and elastic constants are reasonable but deviate somewhat
from the DFT
reference (Table \ref{tab:diamond}). It was shown previously that a GAP
model trained for the
crystalline phases exclusively can reproduce the benchmark even better; \cite{Bartok2010}
here, the gain in transferability (being able to model amorphous as well as crystalline phases)
comes at a small price in terms of accuracy. 

\begin{table}[tb]
\centering
\caption{Structural and elastic properties of diamond,
computed using DFT-LDA and our GAP 
as well as the screened Tersoff potential from
Ref.\ \citenum{Pastewka2013} (``scrT'').}
\label{tab:diamond}
\begin{tabular}{lccc}
\hline
\hline
 & DFT  & GAP & scrT  \\
\hline
$a$ (\AA{})     &  3.532 & 3.539 & 3.566 \\
\hline
$B_{\rm Voigt}$ (GPa) & 466 & 438 & 427 \\
\hline 
$C_{11}$ (GPa)  &  1101  & 1090 & 1073 \\
$C_{12}$ (GPa)  &  148   & 112  & 104 \\
$C_{44}$ (GPa)  &  592   & 594  & 640 \\
\hline
\hline
\end{tabular}
\end{table}

Similar tests for the graphite $c$ parameter gave
6.625 \AA{} (DFT) and 6.518 \AA{} (GAP). 
Despite this slight overbinding (--1.6\%), the agreement is
appreciable, especially given that the Tersoff and Brenner potentials are  short-ranged and 
cannot describe the interlayer spacing in graphite at all
($r_{\rm cut} < c/2$). 

\begin{figure}[tb]
\centering
\includegraphics[width=\figurewidth]{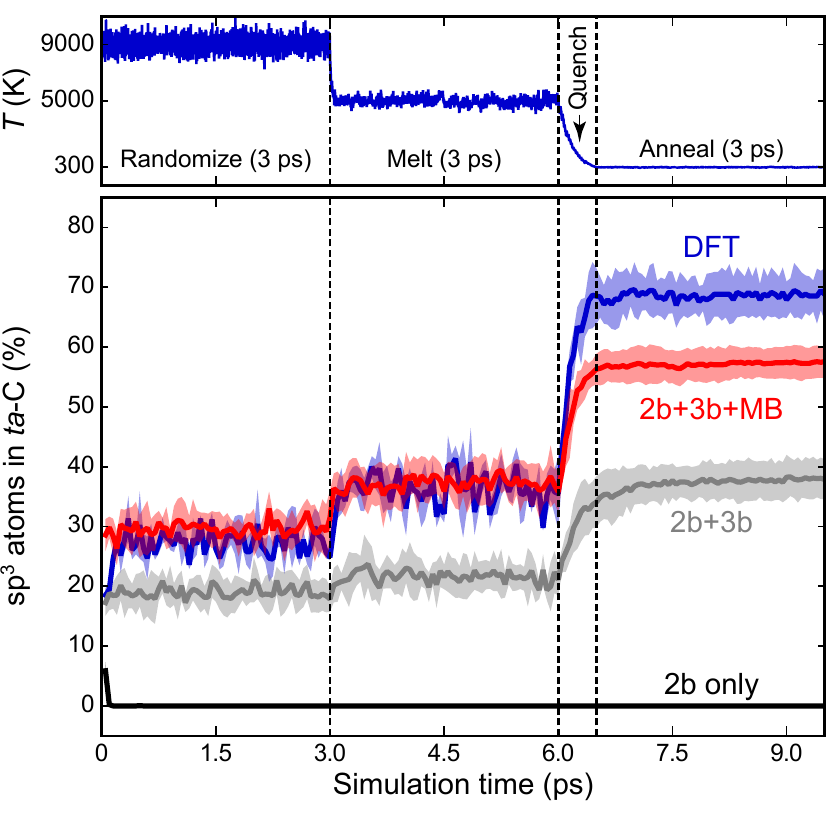}
\caption{\label{fig:sp3_over_time}
         {\em Top:} Exemplary temperature profile during a DFT-MD 
         melt--quench simulation to yield a 216-atom structure of $ta$-C. 
         {\em Bottom:} Concentration of sp$^{3}$ atoms during these
         cycles, measured by counting atomic neighbors up to a cutoff distance
         of 1.85 \AA{}. DFT benchmarks are compared to GAP results
         using different combinations of descriptors; areas of light
         shading indicate standard deviations.}
\end{figure}

\begin{figure}[tb]
\centering
\includegraphics[width=\figurewidth]{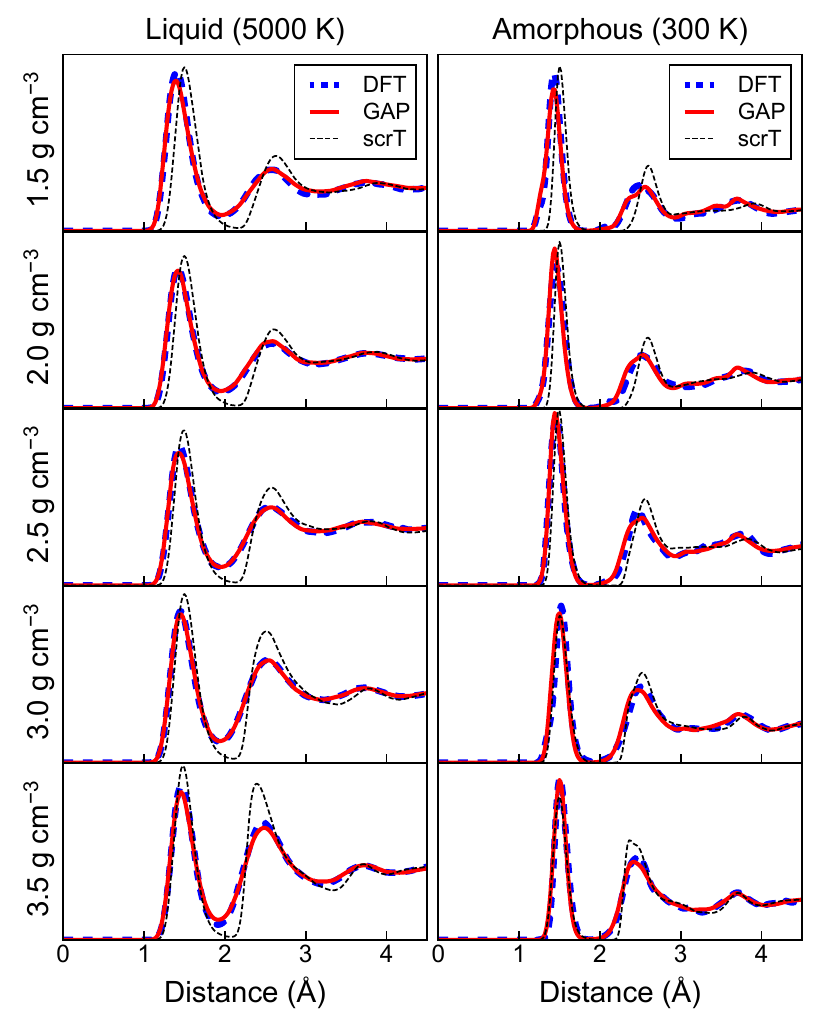}
\caption{\label{fig:rdf}
Radial distribution functions for liquid ({\em left}) and subsequently
quenched amorphous ({\em right}) carbon structures (ten independent 216-atom
structures were created at each density). Results for five densities are
given, spanning the entire range visualized in Fig.\ \ref{fig:structures}.
``scrT'' denotes the screened Tersoff potential as introduced in
Ref.\ \citenum{Pastewka2013}.}
\end{figure}

We now turn to the main subject of the present work: the liquid and amorphous phases
of carbon. We begin by inspecting the concentration of sp$^{3}$ atoms during 
melt--quench simulations (Fig.\ \ref{fig:sp3_over_time}), and use this to once
more assess the performance of different combined structural descriptors. The
DFT reference (blue) shows that in liquid carbon at 3.0 g cm$^{-3}$, approximately
one-third of the atoms are in fourfold coordination, and this number increases
strongly when quenching (6.0$\rightarrow$6.5 ps in simulation time). 
During annealing at 300 K, the DFT average then levels
off at $\approx 70$ \%; as only three structures were created with DFT, fluctuations
and standard deviation (light blue shading) are sizeable. The GAP results, by
comparison, clearly identify the need for combined structural descriptors when
aiming to make physically meaningful predictions: using the two- and three-body
descriptors only yields systematically too low sp$^{3}$ concentrations (gray),
whereas both combined with SOAP essentially reproduce the DFT data for the
liquid forms (red); the sp$^{3}$ count is still underestimated in the quenched
amorphous phase. We performed additional GAP simulations in which we 
increased the quenching time from 0.5 to 2.0 ps, but this did not further improve the result.  
For completeness, we include in Fig.\ \ref{fig:sp3_over_time} results for a
GAP model that employs two-body descriptors {\em only}---but in this case the atoms 
clump  into  unphysical structures within the first few steps  (black line),
not unexpectedly so.

\begin{figure}[tb]
\centering
\includegraphics[width=\figurewidth]{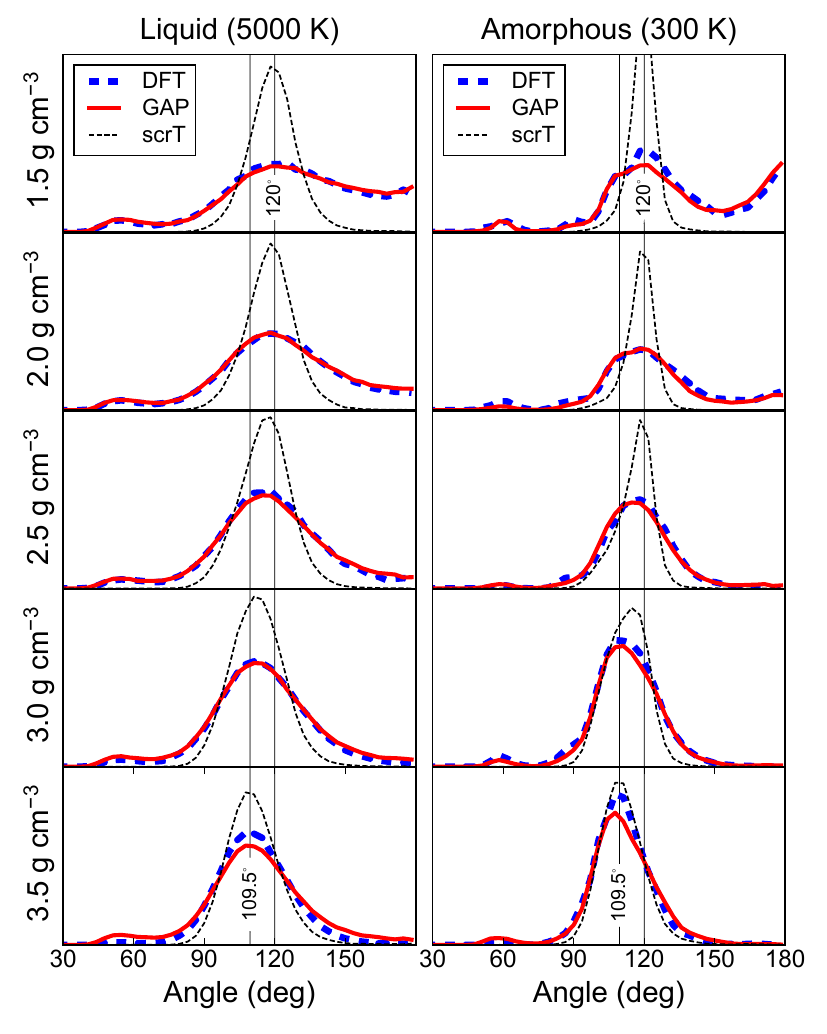}
\caption{\label{fig:adf}
As Fig.\ \ref{fig:rdf}, but for angular distribution functions.
Thin vertical lines mark angles of 109.5$^{\circ}$ and 120$^{\circ}$, corresponding
to ideal sp$^{3}$ (diamond-like) and sp$^{2}$ (graphite-like) motifs,
respectively. The scaling of the vertical axes is arbitrary.}
\end{figure}

The simplest measure of short-range  order in a liquid or amorphous
structure is given by the radial distribution function (RDF). In Fig.\ \ref{fig:rdf},
we compare GAP results to those of DFT,
and start by noting that both are very close. The liquid structures are more diffuse and
less strongly ordered, and the RDFs show a  nonzero first minimum at $\approx 1.9$
\AA{}, whereas the amorphous structures exhibit a gap between their first
and second RDF peak. A small but visible asymmetry of the
second RDF peak at $\approx 2$ \AA{} for all amorphous structures
indicates the presence of fourfold
rings. The screened Tersoff potential (``scrT'') does not predict
the existence of such fourfold rings in $a$-C 
(we will return to this below),
and other than DFT and GAP it lowers the first RDF minimum to almost zero
in all liquid structures.

Figure \ref{fig:adf} shows  angular distribution functions (ADFs).
The ADF maxima at low (high) density are centered around 120$^{\circ}$ 
(109$^{\circ}$), respectively, loosely mirroring the defining structural
features of the crystalline allotropes (graphite honeycombs and diamond
tetrahedra); naturally, this distribution is broader in the highly diffuse
liquids than in the quenched amorphous structures. 
At low densities, a contribution close to 180$^{\circ}$
is seen in the DFT reference data, due to nearly linearly coordinated ``sp''
carbon atoms (yellow in Fig.\ \ref{fig:structures}); this is a minor
feature at 2.5 and 2.0 g cm$^{-3}$, but becomes prominent at
1.5 g cm$^{-3}$, especially so in the quenched amorphous structures
(top right panel in Fig.\ \ref{fig:adf}).
The GAP reproduces these features very well,  both
the location and the extent of the maxima, as well as the overall shape of the
ADF curves. 
The screened Tersoff potential deviates significantly from the DFT
and GAP results, and the ADFs derived from it are zero both at $60^{\circ}$
(absence of threefold rings) and at $180^{\circ}$ (absence of linear
``sp-bonded'' chains).

\begin{figure}[tb]
\centering
\includegraphics[width=\figurewidth]{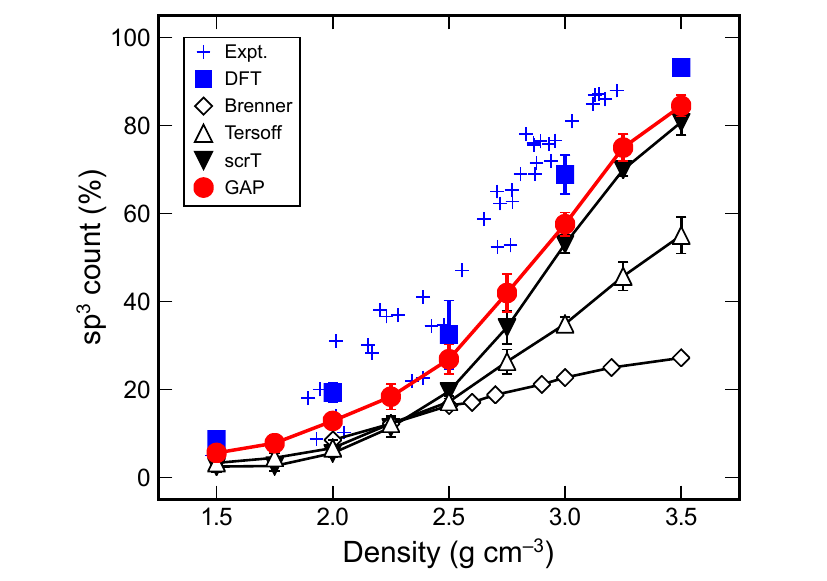}
\caption{\label{fig:sp3_ratio}
Count of ``sp$^{3}$'' (fourfold coordinated) carbon
atoms in quenched $a$-C structures as a function of density.
Ten independent melt--quench cycles were performed at each
density for the empirical and GAP models; three
independent ones were done for DFT.
Data for the Brenner potential are taken from Ref.\ \citenum{Pastewka2008}.
Experimental data have been collected from Refs.\ \citenum{Fallon1993,
Schwan1996,
Ferrari2000}.
Error bars represent standard deviations.
Lines between data points are only guides to the eye.}
\end{figure}

\subsection{Coordination statistics and medium-range order}

Among the key structural characteristics for $a$-C is the concentration
of fourfold coordinated (``sp$^{3}$'') atoms as function of
the sample density. We assess this in Fig.\ \ref{fig:sp3_ratio},
comparing GAP results to DFT but also to previous modeling and 
experimental studies. The empirical Brenner and Tersoff potentials,
as is known,\cite{Pastewka2008} underestimate the sp$^{3}$ count at high density;
indeed, one of the breakthrough successes of the 
screened Brenner and Tersoff potentials has been their much improved
description of $ta$-C in this respect. \cite{Pastewka2008}
In comparison, the GAP data (red in Fig.\ \ref{fig:sp3_ratio}) are even closer
to the DFT reference (blue), particularly at lower densities.
The residual error of the GAP results
is most pronounced at 3.0 g cm$^{-3}$, and so using this density for the  example in
Fig.\ \ref{fig:sp3_over_time} showed the worst of all cases.

\begin{figure}[tb]
\centering
\includegraphics[width=\figurewidth]{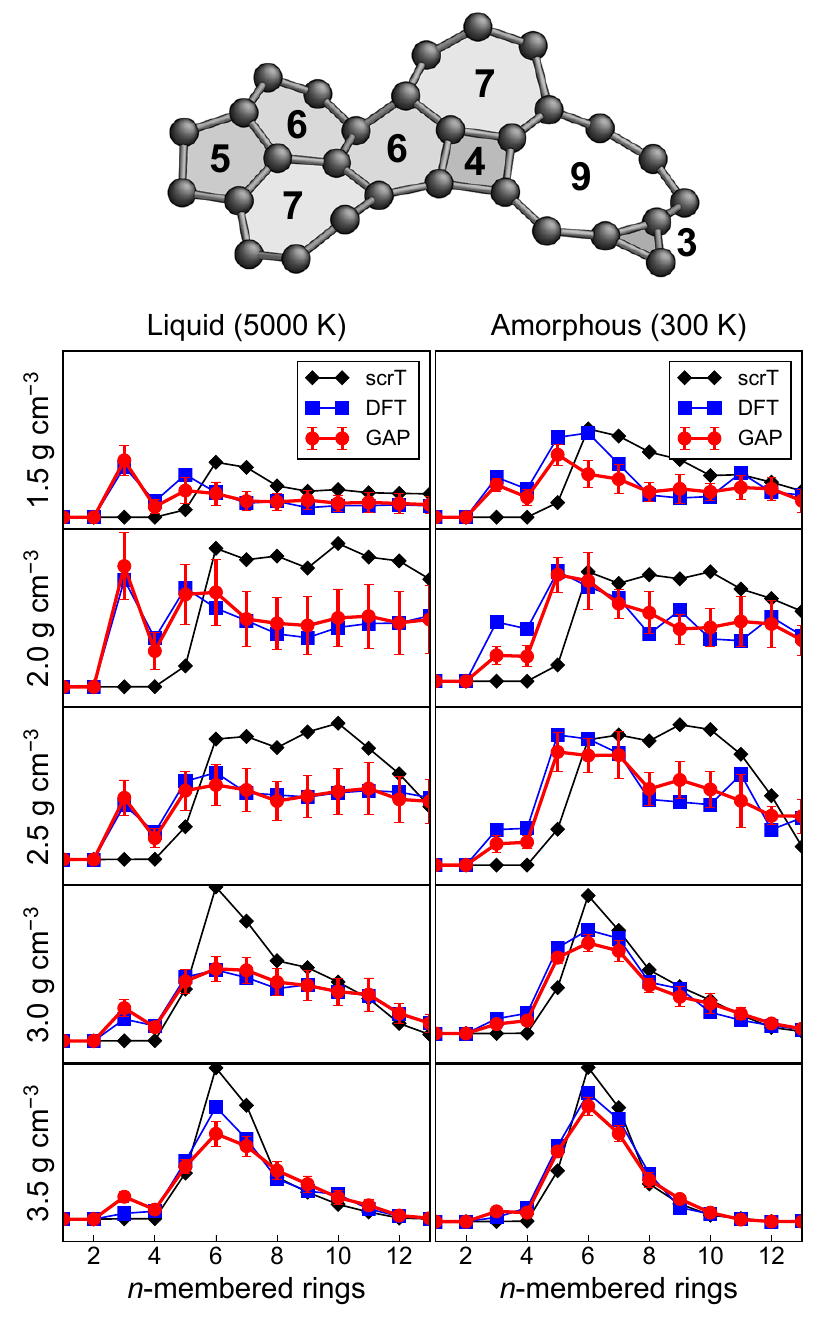}
\caption{\label{fig:rings}
Medium-range order in $a$-C as evaluated through ring statistics.
{\em Top:} Structural fragment from one of the DFT generated $a$-C structures
at 2.0 g cm$^{-3}$,  chosen  to visualize the diversity of
 ring sizes observed. Rings are indicated by shading, and their
 size $n$ is given. 
{\em Bottom:} Ring statistics for liquid ({\em left}) and quenched
amorphous ({\em right}) carbon structures obtained from DFT (blue), GAP
(red), and screened Tersoff potential (``scrT''; black) simulations.
Data for the liquid structures have been collected over the last 1 ps
of the respective trajectory; data for the amorphous structures correspond
to the last snapshot only, as the structures are strongly correlated in this
case. For GAP-derived structures, the standard deviation
for the count of each ring size is indicated by error bars.}
\end{figure}

Looking beyond the first nearest-neighbor shell, the medium-range order 
in amorphous materials is conventionally
characterized by means of ring statistics, which we evaluate using
Franzblau's shortest-path algorithm.\cite{Franzblau1991} Again, we
compare liquid and quenched amorphous structures side-by-side, and
inspect the entire range from low to high densities (Fig.\ \ref{fig:rings}). 

The DFT reference (blue) shows that the distribution is quite complex:
at high densities, the ring sizes center around
sixfold (similar to diamond, where $n=6$ exclusively), and the distribution decays
quickly beyond that; no large-membered rings are found
in $ta$-C. By contrast, the distribution in the low-density structures
is less clearly defined and does involve higher-order rings, indicative
of structural voids. 

The results for $ta$-C at
300 K are very similar with all three methods. In addition, the GAP model also recovers
the three- and four-membered rings that are key features of the liquid and
also prominent  in low-density amorphous structures.\cite{Galli1989}
The screened Tersoff potential, by contrast, overestimates the average
ring size in low-density $a$-C, and does not predict the occurrence
of any three- or fourfold rings, neither in the liquid nor in the amorphous
phases.

\begin{figure}[tb]
\centering
\includegraphics[width=\figurewidth]{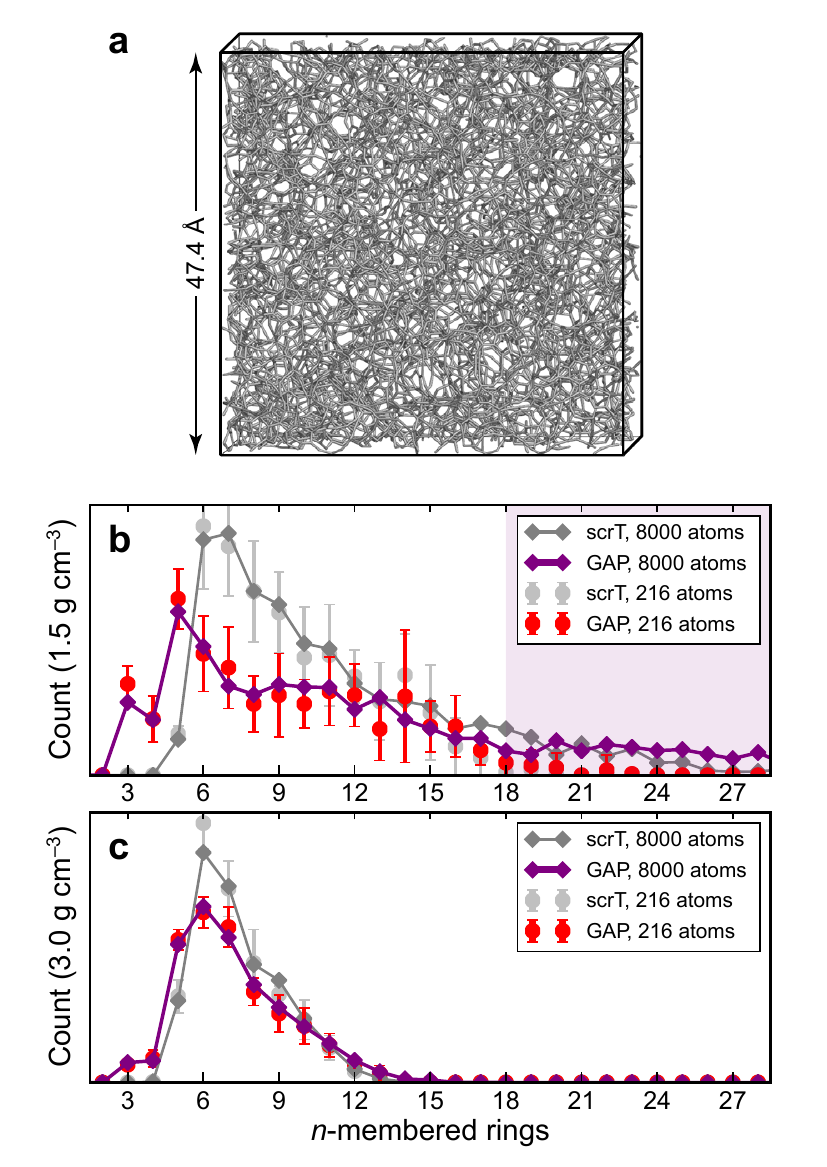}
\caption{\label{fig:bulk_8000}
Application of the GAP to larger-scale simulations. 
(a) Melt-quenched 8,000-atom structure of $a$-C at 1.5 g cm$^{-3}$, shown as
stick drawing.
(b) Ring statistics for this structure (purple) and as averaged over 10 different 216-atom
structures (red; as in Fig.\ \ref{fig:rings}). Purple shading
emphasizes
ring sizes of $n \geq 18$ that the smaller systems cannot reproduce.
(c) Same analysis but for $ta$-C (3.0 g cm$^{-3}$). }
\end{figure}

So far, all validation of the new potential has been done against DFT, and therefore
necessarily been limited to rather modest system sizes of 216 atoms.
The true strength of ML potentials, however, is in the application to
larger structures.
Figure \ref{fig:bulk_8000} shows results for an 8,000-atom $a$-C structure
at 1.5 g cm$^{-3}$, which would  presently be impossible to generate with 
DFT-based MD even on state-of-the-art supercomputers. 

RDFs and ADFs obtained with 216-atom 
and 8,000-atom structures are practically the same and are hence not shown.
The situation for the
ring statistics (Fig.\ \ref{fig:bulk_8000}b) is more complex. For small- and
medium-sized rings, 
results for the large system (purple) come very close to the
{\em average} from the 216-atom structures (red). Hence,
while a single 216-atom snapshot will not be sufficient to investigate ring
statistics of $a$-C models, one may instead collect averages over
sufficiently many smaller structures, and therefore reproduce the short- and 
medium-range structural features without requiring larger simulation cells.
Nonetheless, there {\em is} an inherent deviation
between the 216- and 8000-atom structures: namely, at very large ring sizes
($n \geq 18$; shaded in Fig.\ \ref{fig:bulk_8000}b), which the 216-atom cells
cannot reproduce as they are simply too small.
This emphasizes that realistic studies of voids and porosity in $a$-C will
require large structures on the order of at least several thousand atoms. 

We also created an 8,000-atom $ta$-C structure (Fig.\ \ref{fig:bulk_8000}c): 
in this case, no voids are found but a dense, ``diamond-like''
network. Consequently, no large rings
($n > 15$) are observed, and the 216-atom simulations
already provide a very good estimate of the medium-range structural order.
Likewise, the screened Tersoff potential here correctly reproduces the maximum
at $n=6$ as well as the abundance of larger-membered
rings. The latter drops to zero between $n=12$ and $n=15$ for all potentials and
system sizes investigated.

\begin{figure}[tb]
\centering
\includegraphics[width=\figurewidth]{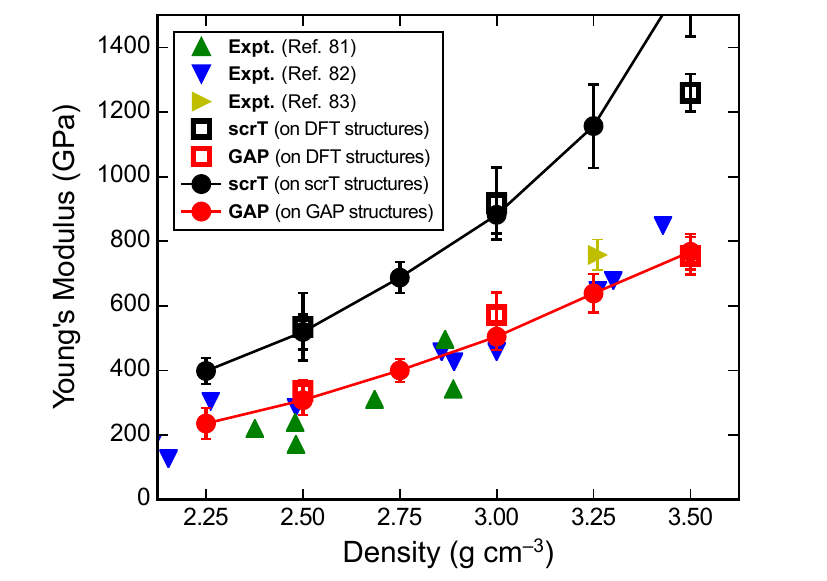}
\caption{\label{fig:elastic_constants}
Young's modulus of $a$-C as a function of density.
Experimental values are taken from 
Ref.\ \citenum{Schultrich1996} (green),
Ref.\ \citenum{Schultrich1998} (blue), and 
Ref.\ \citenum{Ferrari1999a} (yellow), respectively. 
Lines between data points are guides to the eye.}
\end{figure}

\subsection{Elastic properties}

We  next evaluated the Young's modulus of $a$-C which, like the sp$^{3}$
count, depends strongly on density.
\cite{Schultrich1996, Schultrich1998, Ferrari1999a}
We compare results using scrT and GAP, but not DFT, due to the high
expense of fully relaxing the internal degrees of freedom for several
uncorrelated models.
In addition, we  disentangle the effect of
input structure versus potential performance for the
prediction of elastic properties,
and therefore also use both scrT and GAP to evaluate the Young's moduli of 
DFT-generated structures. 

To compute the Young's modulus of $a$-C, we take previously generated 216-atom
structures, perform further short MD quenches from
300 K to very low temperature, and finally a conjugate-gradient relaxation to
minimize the forces on atoms; the cell vectors remain fixed to keep
the density unchanged. For each optimized structure, we compute the
full $6 \times 6$ matrix of elastic constants {\bf C} without imposing symmetry
operations, and invert this matrix to obtain the compliance matrix {\bf S}.
\footnote{J.~F.~Nye, {\em Physical Properties of Crystals: Their Presentation by
Tensors and Matrices} (Clarendon, Oxford, 1969).}
From this, we calculate the Young's modulus $E$ 
(see, e.g., Ref.\ \citenum{Brantley1973})
by averaging over the three spatial directions:
\begin{equation}
 E = \frac{1}{3} \left[ \frac{1}{S_{11}} + \frac{1}{S_{22}} + \frac{1}{S_{33}} \right]
\end{equation}
and subsequently over independent structures (ten from scrT and GAP melt--quench
runs, three for the DFT case; see above). The GAP results agree very well
with experiments at all relevant densities
(Fig.\ \ref{fig:elastic_constants}), and as expected they predict increased
stiffness as density and sp$^{3}$ concentration (``diamond-likeness'')
increase. The screened Tersoff potential correctly captures the same trend,
albeit the absolute values are significantly overestimated; this is most pronounced at
higher densities.

\begin{figure}[tb]
\centering
\includegraphics[width=\figurewidth]{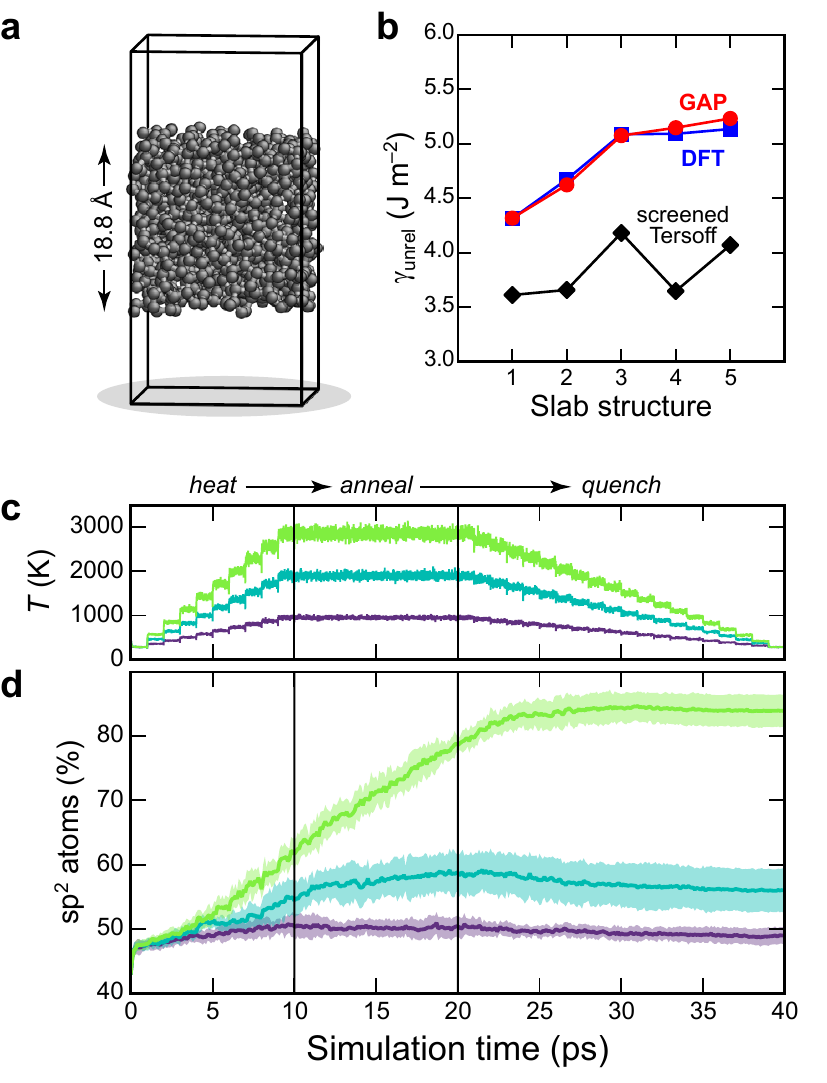}
\caption{\label{fig:surface_reconstructions}
(a) Exemplary surface slab of $ta$-C, freshly cleaved
         from a 1000-atom bulk structure.
(b) Unrelaxed surface energies (Eq.\ \ref{eq:gamma}) for
         five slabs cleaved from the same bulk structure.
         Lines between data points are guides to the eye.
(c) Course of temperatures in the protocol we use to generate reconstructed
surfaces: the systems are heated over 10 ps to 1000 (green), 2000 (yellow),
or 3000 K (red), respectively, and annealed at this temperature for another
10 ps. The final 20 ps then constitute a slower
cooling back to 300 K.
(d) Concentration of sp$^{2}$ carbon atoms in 1000-atom 
slabs versus simulation time.
Averages over ten independent structures are given, and areas
of light shading indicate standard deviations.}
\end{figure}

\begin{figure*}[tb]
\centering
\includegraphics[width=17.8cm]{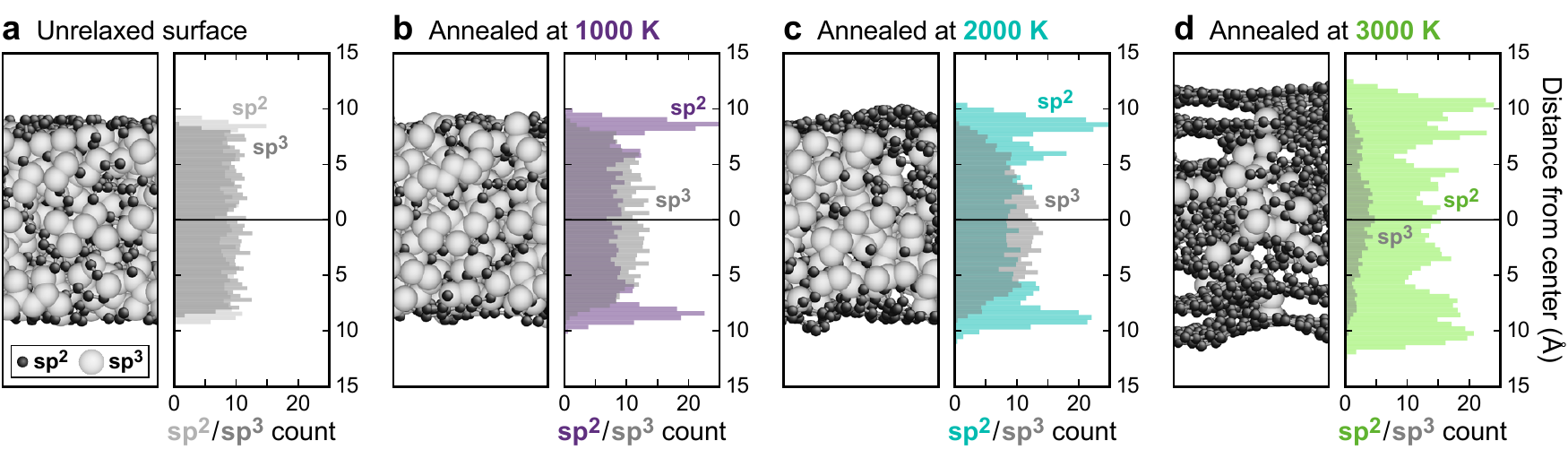}
\caption{\label{fig:surface_profiles}
Surfaces of {\em ta}-C before (a) and after annealing at different temperatures (b--d)
as predicted from GAP simulations. Each panel shows one exemplary slab structure on
the left, with sp$^{2}$ (``graphite-like'') atoms drawn as small dark spheres,
and sp$^{3}$ (``diamond-like'') atoms larger and light gray. Note how annealing
at 3000 K leads to beginning detachment of graphitic-like sheets, and disintegration
of the $ta$-C regions even in the center of the slab. Histograms for the 
sp$^{2}$/sp$^{3}$ distribution in direction normal to the surface have been
collected over ten independent structures; the count is normalized per
simulation cell.}
\end{figure*}

\subsection{From the bulk to surfaces}

Realistic materials modeling, especially at the nanoscale,
must extend from the bulk to a description of crystal
surfaces and their reactivity.\cite{[{}][{, and references 
therein.}]ChalcNC_Review}
Likewise, the surfaces of amorphous matter are of broad interest
but pose particular and significant challenges for modeling. 
We here present initial applications of our GAP to amorphous carbon surfaces
of the 3.0 g cm$^{-3}$ phase ($ta$-C).
This is because dense, diamond-like carbon is used in coatings\cite{Robertson2002}
and it is this form for which surface phenomena are most relevant.

Early studies of $ta$-C surfaces
have been reported at the DFT level but have
necessarily been restricted to very small system sizes.\cite{Dong1998, *Chen2006,
*Bauschlicher2010, *Caro2014}
Larger-scale simulations were made possible by 
tight-binding schemes \cite{Haerle1999} and EDIP,\cite{Marks2006a, *Marks2006}
but even high-quality empirical potentials may face problems when it
comes to the prediction of surface energies; this has already been reported
for diamond. \cite{Pastewka2013}

 Conventionally,
 the surface energy, $\gamma$, is calculated as
\begin{equation}
 \label{eq:gamma}
 \gamma = \frac{1}{2A} \left[ E_{\rm slab} - N \times E_{\rm bulk} \right]
\end{equation}
for an elemental (or stoichiometrically precise) surface slab that contains
$N$ atoms and exposes equivalent surface areas $A$ at top and bottom;
in this expression, $E_\text{slab}$ denotes computed total energies 
for a slab model per unit cell, and $E_\text{bulk}$ refers to the energy
of the underlying bulk structure per atom.

For amorphous systems, the structure of the surface is not uniquely defined
(there are no distinct cleavage planes as in crystals),
and to calculate $\gamma$ one must average over many large structures.
We assess the suitability of the GAP\ model for such studies
by computing surface energies of {\em ta}-C and comparing to DFT values.
We used a GAP to generate a 1000-atom  bulk {\em ta}-C structure and
cleaved five different surfaces normal to the [001] direction of
the simulation cell (cf.\ Fig.\ \ref{fig:surface_reconstructions}a).
For each surface, the unrelaxed surface energy was  
evaluated using the three methods  
(Fig.\ \ref{fig:surface_reconstructions}b). 
The GAP model fully reproduces the stability ordering; for
the most stable surface (structure 1), GAP and DFT results differ by 
less than 0.01 J m$^{-2}$.
For the two least stable candidates, 4 and 5, this difference increases
slightly but remains small (below 0.1 J m$^{-2}$, or 2\%). The screened
Tersoff potential yields much lower surface energies, similar to what
has been reported for diamond.\cite{Pastewka2013} 

We finally perform high-temperature annealing simulations with our GAP,
to assess structural relaxations and reconstructions at $ta$-C surfaces.
These are associated with an increased formation of sp$^{2}$ atoms
(``graphitization'') that has been observed in several {\em ex-situ} experiments 
\cite{Friedmann1996, *Sullivan1997, *Ferrari1999}
and also {\em in situ} during film growth. \cite{Chhowalla1997}
A discussion of the relevant differences between experiment and theory
has been given by Marks. \cite{Marks2006a, *Marks2006}
Higher temperatures than in experiment must be used to overcome kinetic
barriers during simulation---as experiments typically involve up to 
one hour of annealing.\cite{Ferrari1999} In that sense, the {\em absolute}
annealing temperature used for simulation is fictious;\cite{Marks2006a, *Marks2006} 
its choice depends on the computational method,\cite{deTomas2016} and a
suitable annealing temperature must therefore be found by trial and error.

In Fig.\ \ref{fig:surface_reconstructions}c--d we explore the use of different
such temperatures, and in particular we analyze the structures obtained before
and after each of the different annealing runs (Fig.\ \ref{fig:surface_profiles}).
In each simulation, the slab is gradually heated to
the target temperature over 10 ps, annealed for 10 ps,
and then cooled back to 300 K over another
20 ps; each structure contains 1000 atoms, and ten independent ones are studied
in parallel to improve statistics. Monitoring the concentration of sp$^{2}$ atoms
during these simulations provides the most direct insight:
heating to 1000 K induces no significant
changes, albeit it ``heals'' the dangling bonds directly at
the surface (indicated by a small jump in sp$^{2}$ immediately upon initializing the
simulation); therefore, the 1000 K annealed structure may be a useful
representative of the {\em non}-graphitized surface. 
At the  intermediate setting of 2000 K,  the sp$^{2}$ concentration in the system
rises slightly during annealing and is then lowered again during cooling;
the interior of the slab and its density 
remain close to that of bulk $ta$-C, whereas reconstructions are observed
at the surface. Finally, heating to 3000 K
graphitizes the entire system; this is reminiscent of what was seen earlier
by Powles and co-workers using the EDIP model.\cite{Powles2009} It also
leads to a
strong expansion of the slab interior (Fig.\  \ref{fig:surface_profiles}d).

\begin{figure}[tb]
\centering
\includegraphics[width=\figurewidth]{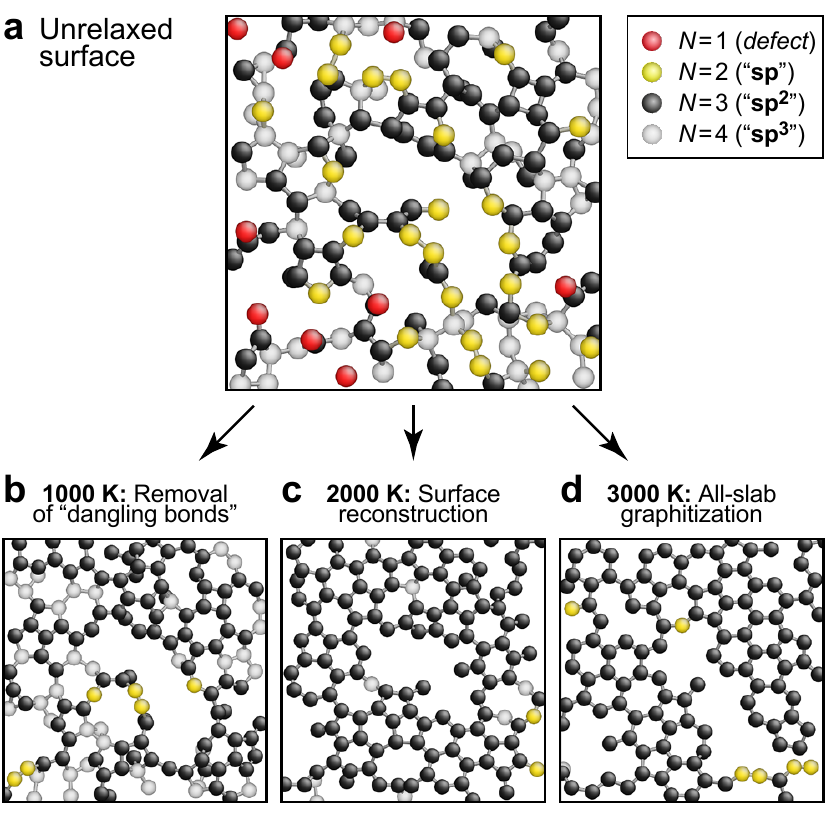}
\caption{\label{fig:surfaces_topview}
Top views of the same surface structure before (a) and after (b--d) different
degrees of annealing. Only atoms in the outermost 3 \AA{} are shown, and
coloring indicates the coordination number.}
\end{figure}

A top view best visualizes the atomic-scale processes at the surface
(Fig.\ \ref{fig:surfaces_topview}). The freshly cleaved, unrelaxed structure
shows a number of ``dangling bonds'' and low-coordinated atoms, trivially so
as the tetrahedra in $ta$-C have been cut apart. These defects largely disappear
during annealing at 1000 K already, but at this temperature the
surface stays strongly disordered (Fig.\ \ref{fig:surfaces_topview}b).
By contrast, increasing the annealing temperature to 2000 K leads to 
graphitized layers of several \AA{} thickness at the surfaces 
(Fig.\ \ref{fig:surface_profiles}c): sixfold rings are seen, as 
well as pairs of five- and sevenfold ones that are likely metastable
(Fig.\ \ref{fig:surfaces_topview}c); still, the surface atoms are connected
to lower-lying sp$^{3}$ atoms even within the topmost 3 \AA{}, and the graphitization
therefore remains a genuine surface phenomenon.
By contrast, during annealing at 3000 K, the entire
slab graphitizes as seen above, and a strongly defective graphene sheet begins
to detach from the surface; no near-surface sp$^{3}$ atoms are seen any more
(Fig.\  \ref{fig:surfaces_topview}d).

\section{Conclusions}

We have developed a machine-learning based GAP model for atomistic simulations of
liquid and amorphous
elemental carbon. 
The structural complexity that the potential has to encompass,
as well as the nonlocality in forces are notable,
and  larger than in any previous ML-based interatomic potential model.
Nonetheless, our GAP predicts energies that are
largely in the range of tens of meV/atom; 
characteristic structural properties, such as the sp$^{3}$
count and the medium-range order as expressed through ring statistics,
are faithfully recovered, and surface energies and reconstructions are well
described by the GAP. 

The central issue in the development of atomistic materials modeling remains 
in the tradeoff between accuracy and cost. 
The GAP model presented here is many orders of magnitude faster  than
DFT, but  slower than state-of-the-art empirical
potentials (while similarly linear scaling).
Being thus intermediate between both realms, GAP models appear to be promising
tools for accurate large-scale atomistic simulations, including amorphous
materials and their surfaces.\\     

\begin{acknowledgments}
We thank L.\ Pastewka, A.\ P.\ Bart\'o{}k, J.\ R.\ Kermode, and 
S.\ R.\ Elliott for ongoing valuable discussions and for helpful
remarks on this work.
V.L.D.\ gratefully acknowledges a postdoctoral fellowship
from the Alexander von Humboldt Foundation, as well as support
from the Isaac Newton Trust (Trinity College Cambridge). 
This work used the ARCHER UK National Supercomputing
Service (http://www.archer.ac.uk) via 
EPSRC Grant EP/K014560/1. 
\end{acknowledgments}

\cleardoublepage

\onecolumngrid
\begin{center}
\textbf{\large Supplementary Information}\\[12mm]
\end{center}
\twocolumngrid

\setcounter{figure}{0}
\setcounter{table}{0}
\setcounter{page}{1}
\makeatletter
\renewcommand{\thefigure}{S\arabic{figure}}
\renewcommand{\bibnumfmt}[1]{[S#1]}
\renewcommand{\citenumfont}[1]{S#1}
\renewcommand{\thepage}{S\arabic{page}}
\renewcommand{\thetable}{S\Roman{table}}

\section*{Composition of training database}

The database is kept in extended {\tt xyz} format, which allows us to store
data for atomic positions and forces, as well as total energies and virials,
all in the same file. Individual configurations are concatenated, and identified
by two custom keywords, {\tt config\_type} and {\tt detailed\_ct}, respectively.
The first is used for grouping the data during training. The second, more
detailed configuration type indicates how exactly the training data have 
been obtained. A description is provided in the following.

For all parts of the dataset, the number of points given here refers to the
structures that in the (randomly sampled) {\em training set},
which comprised 90\% of the total database. For example, for the
``{\tt step0}'' dataset, a total of 500 structures were computed, and from these
a~random number generator selected 462 entries for the training and 38 entries
for the test set.\\

\small

\noindent
{\bf Bulk liquid and amorphous carbon structures ({\tt config\_type=bulk\_amo})}

\begin{itemize} 
 \item {\tt detailed\_ct=step0} (462 points, 64 at./cell): Liquid carbon at 9000 K; five
 {\tt cp2k} trajectories at densities of 1.5, 2.0, 2.5, 3.0, and 3.5 g cm$^{-3}$,
 respectively. Each trajectory had 3000 steps, from which every 30 steps
 a snapshot was taken and added to the database.
 \item {\tt detailed\_ct=LDL} (166 points, 64 at./cell): Low-density liquid: additional
 {\tt cp2k} snapshots at densities of 1.50 and 2.00 g cm$^{-3}$, taken both at
 9000 K and 5000 K. These were added at an intermediate 
 stage during training, to prevent the GAP from predicting too many ``sp'' chains
 in the low-density liquids.
 \item {\tt detailed\_ct=qu\_cp2k} (88 points, 64 at./cell): Liquid structures created with a
 preliminary GAP (trained on {\tt step0}) that were then equilibrated at 5000 K
 (0.5 ps) and quenched (0.5 ps) to 300 K using {\tt cp2k}. 
 Ten uncorrelated structures were generated at 2.0 cm$^{-3}$, and ten at 3.0 cm$^{-3}$.
 \item {\tt detailed\_ct=qu\_tetra} (107 points, 64 at./cell): Melt--quench trajecories
 with a focus on higher
 densities (richer in tetrahedral motifs). We repeated the above for high-density structures,
 generating ten structures each at 3.25 and 3.5 g/cm$^{-3}$, respectively; this was done
 to include structures more rich in tetrahedra early on. 
 \item Iterative training: Iteration 2
   \begin{itemize}
   \item {\tt detailed\_ct=iter2\_1} (274 points, 27--64 at./cell): Using an intermediate
   GAP (with distance, angle, and SOAP descriptors, the latter employing 1000 sparse
   points; hence denoted {\tt das1000}), trained using a part of the above data, 
   100 independent melt--quench trajectories were performed
   with randomized densities between 1.5 and 3.5 g cm$^{-3}$. From each, three structures
   were extracted: one from the equilibrated 5000 K liquid (at 6.00 ps), one during the
   quench (at 6.15 ps), and one final step representing the equilibrated amorphous phase
   (at 9.50 ps). For each, a single-point computation was performed and added to the
   database.
   \item {\tt detailed\_ct=iter2\_2} (86 points, 27--64 at./cell): Same but using a
   GAP model with a larger number of SOAP sparse points ({\tt das2000}); 100 independent
   trajectories from each of which only the
   final step was taken. 
   \item {\tt detailed\_ct=iter2\_3} (186 points, 64 at./cell): As before but using
   a smaller GAP model ({\tt das0250}); 200 independent melt--quench trajectories from
   each of which only the final step was taken.
   \item {\tt detailed\_ct=iter2\_4} (272 points, 64--125 at./cell): Starting structures
   increased to include $4 \times 4 \times 5$ (80 at.), $4 \times 5 \times 5$ (100 at.),
   and $5 \times 5 \times 5$ (125 at.) expansions.
   \end{itemize}
 \item Iterative training: Iteration 3
    \begin{itemize}
   \item {\tt detailed\_ct=iter3\_1} (260 points, 27--64 at./cell): Using the previous
   data including the second iteration, a new GAP model was fitted and this was in turn
   used to generate structures as described for {\tt iter2\_1}.
   \item {\tt detailed\_ct=iter3\_2} (87 points, 27--64 at./cell): As before, but with
   a GAP model using slightly modified parameters compared to {\tt iter3\_1};
   in this case, only the last step from each trajectory was used.
   \end{itemize}
 \item Iterative training: Iteration 4
    \begin{itemize}
   \item {\tt detailed\_ct=iter4\_1} (272 points, 27--64 at./cell): Iterative training as 
   before, generating structures as described for {\tt iter2\_1}.
   \item {\tt detailed\_ct=iter4\_2} (356 points, 27--64 at./cell): Iterative training
   as before,
   generating structures as described for {\tt iter2\_1}, with slightly modified GAP
   parameters compared to {\tt iter4\_1}. 
   \end{itemize}  
 \item Iterative training: Iteration 5
    \begin{itemize}
   \item {\tt detailed\_ct=iter5\_1} (178 points, 125 at./cell): Iterative training
   as before,
   now increasing the system size; 100 independent melt--quench runs were performed,
   from which one snapshot from the 9000 K liquid and one from the amorphous structure
   was taken. 
   \item {\tt detailed\_ct=iter5\_2} (178 points, 125 at./cell): As before, but using
   one snapshot from the 5000 K liquid and one during quenching. In total, four
   structures were thus extracted from each 125-atom melt--quench trajectory.
   \end{itemize}  
 \item {\tt detailed\_ct=iter6\_1} (91 points, 64 at./cell): Iterative training as
 before. In this case, once the amorphous structures had been generated, the lattice
 parameters of the simulation cells were scaled uniformly by a randomized factor
 between 0.95 and 1.02 (to sample structures further away from
 equilibrium);
 subsequently, the atomic positions were relaxed using a 
 conjugate-gradient scheme, and both scaled unrelaxed and scaled relaxed structures
 were added to the training database. 
 \item {\tt detailed\_ct=iter7\_1} (88 points, 64 at./cell): Iterative training as
 before. In this case, amorphous structures were generated and relaxed, and then one
 of the lattice parameters was scaled by a randomized factor between 0.95 and 1.05;
 this strained cell was then added to the database.
\end{itemize}

\noindent
{\bf Amorphous surfaces ({\tt config\_type=surf\_amo})}

\begin{itemize} 
 \item {\tt detailed\_ct=asurf\_cp2k\_from5k} (274 points, 64 at./cell):
 Two bulk liquid structures
 generated with {\tt cp2k} (64 atoms each) were cleaved at five equidistant planes
 to generate ten slabs in total; each of the latter was subjected to 3 ps of
 {\tt cp2k} MD at 1000, 2000, and 5000 K in parallel. This leads to highly disordered
 surface structures with beginning disintegration, and therefore allows us to sample
 a large diversity of near-surface structures. In total, 30 runs were hence performed,
 from which ten structures (every 300 steps) were extracted and added
 to the training database. 
 \item {\tt detailed\_ct=surf\_from\_iter4\_216} (81 points, 216 at./ cell): Bulk
 216-atom {\em ta}-C structures (3.0 g cm$^{-3}$) were created, and subsequently
 cleaved to yield surface slabs; the latter (but not the bulk structures) were
 added to the database.
 \item {\tt detailed\_ct=surf\_unrel\_iter4} (92 points, 64 at./ cell): Bulk 64-atom
 {\em ta}-C structures were cleaved normal to [001], and these slab structures were
 added to the database.  
 \item {\tt detailed\_ct=surf\_unrel\_iter4\_x} (91 points, 64 at./ cell): Same but
 with cleavage normal to [100]. 
 \item {\tt detailed\_ct=surf\_unrel\_iter4\_y} (86 points, 64 at./ cell): Same but
 with cleavage normal to [010].
\end{itemize}

\noindent
{\bf Bulk crystal structures ({\tt config\_type=bulk\_cryst})}

\begin{itemize} 
 \item {\tt detailed\_ct=dist\_dia} (182 points, 8 at./cell): Distorted configurations
 starting
 from the DFT-optimized structure of diamond. We generated 50 structures each with
 all lattice parameters scaled by factors of 0.94, 0.97, 1.00, or 1.03.
 For each cell, none, one, or more of the lattice parameters
 were randomly changed by $\pm 5\%$, and none, one, or more of the angles was randomly
 changed by $\pm 5$ degrees; the other parameters were kept fixed. The atomic positions
 in each cell were randomly either (i) fixed, (ii) displaced randomly by $\pm 0.02$ \AA{},
 or (iii) relaxed using the Tersoff potential. 
 \item {\tt detailed\_ct=dist\_graphite} (174 points): Same as above, but for graphite.
\end{itemize}

\noindent
{\bf Isolated dimer ({\tt config\_type=cluster})}

\begin{itemize} 
 \item {\tt detailed\_ct=dimer} (30 points, 4 at./cell): DFT computations for C$_{2}$
 dimers with bond lengths of 0.8--3.7 \AA{} (in increments of 0.1 \AA{}). In this case, a 
 large cell of $20 \times 15 \times 15$ \AA{}$^{3}$ was used, and reciprocal space
 was sampled at $\Gamma$.
\end{itemize}

\normalsize

\section*{Errors for test versus training set}

\begin{figure}[htb]
\centering
\includegraphics[width=\figurewidth]{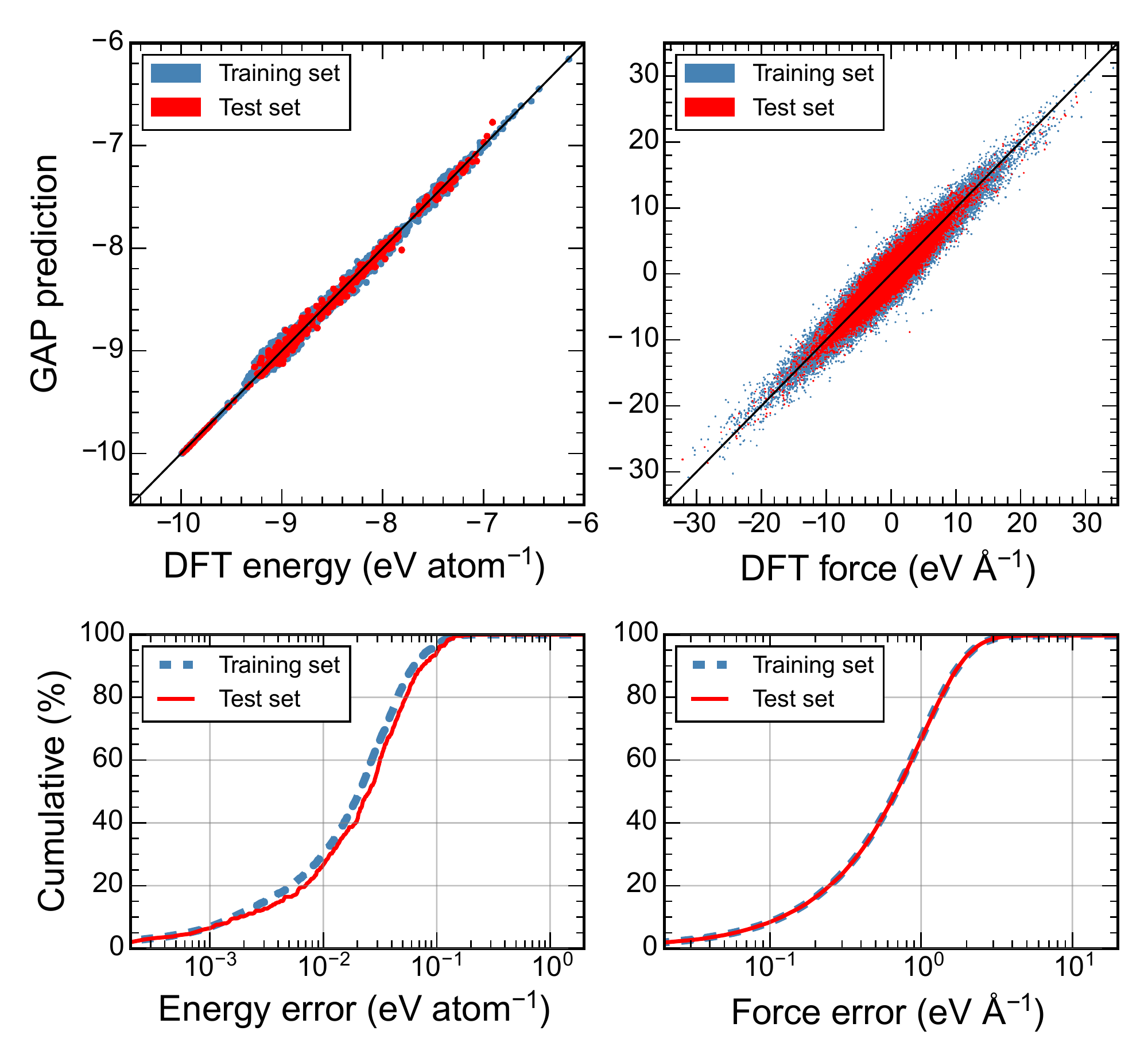}
\caption{\label{fig:scatter_train_test}
Scatterplot of DFT-computed and GAP-predicted total energies and force components
for all 4050 structural models in the training set ({\em blue}) and for a separate test set
of 450 structures ({\em red}) not included in the training. Below, the cumulative error
distributions are given, which are very close for the energies, and 
practically indistinguishable for the force components.}
\end{figure}

To test the performance of an ML potential outside its training range,
it has been advocated to take a certain subset from the structural database
as ``test set'' that is not included in the training. Here, doing so
evidences that the errors in both subsets are practically superimposable
(Fig.\ \ref{fig:scatter_train_test}): this is encouraging with regard to the quality
of the potential, but also to the completeness of the underlying database, as
the test set is randomly drawn from the training data and hence structurally similar
to them. In other words, Fig.\ \ref{fig:scatter_train_test} addresses not
the extrapolation but the {\em interpolation} behavior of our GAP.

\newpage

\section*{Trimer potentials}

Albeit the GAP model reported here is trained for bulk and surface
(that is, extended) structures, we performed additional tests for 
isolated trimers to assess the extrapolation behavior of the GAP.
This case study allows us to more clearly pinpoint the effect of
different training data, and the large improvement in potential
quality that can be achieved by including dimer data during the fit.

Therefore, we place two carbon atoms in a large box, at a
fixed distance of $d({\rm C-C})$ = 1.5, 2.0, or 2.5 \AA{}.
We then place a third atom in the box on finely
meshed grid points; the energy of each system is then evaluated as  
\begin{equation}
  \Delta E = E({\rm trimer}) - E({\rm dimer}) - E({\rm atom}).
\end{equation}

\begin{figure}[b]
\centering
\includegraphics[width=17.8cm]{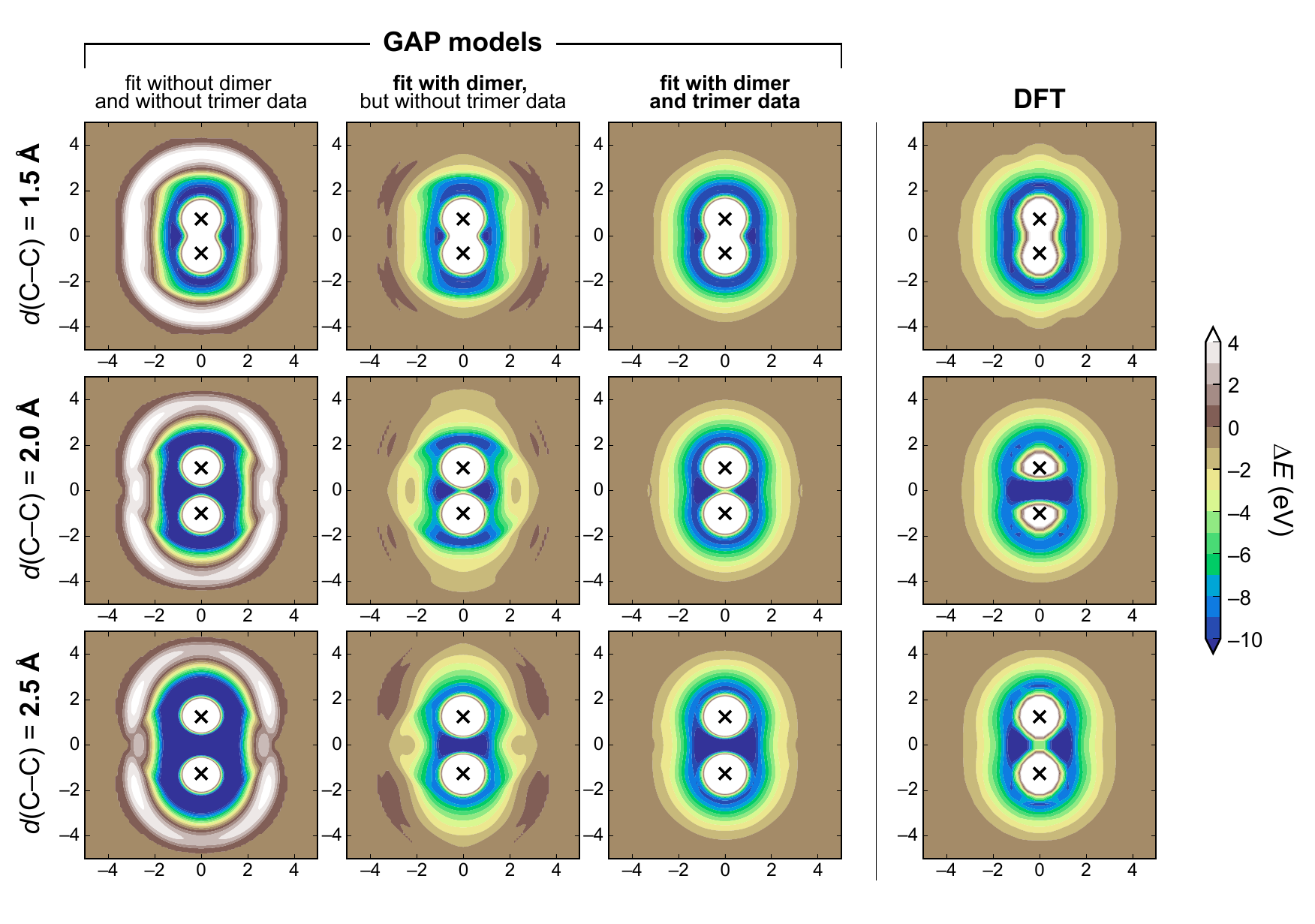}
\caption{\label{fig:SI_trimer_scans}
Potential-energy scans for isolated carbon trimers, computed using different GAP
models (left-hand side) and DFT (right-hand side).
Two carbon atoms (marked as $\times$) are placed in the center
of a simulation box, and a third atom is placed on finely meshed grid points
in the surrounding; at each point, the energy is computed relative to a free
``$\times\;\times$'' dimer at the given spacing plus a free atom. (In other
words, as soon as the third atom is outside the cutoff radius for both central
carbon atoms, the energy drops to zero.)}
\end{figure}

These interaction energies are characterized in Fig.\ \ref{fig:SI_trimer_scans}
and compared to DFT reference data. A GAP model for which the fit did not include
any dimer or trimer data (leftmost column) shows erroneous extrapolation, including an
energy barrier at intermediate distances; no such barrier
is seen in DFT. A GAP model fitted to dimer but not to trimer training data 
reproduces the trimer potential-energy surface qualitatively very well; however,
small deviations remain (middle column). We finally tested the performance of a
model where the DFT datapoints for the trimer
{\em have} been included during training: not surprisingly, this improves
the description of the trimer further. 

\onecolumngrid
\noindent\rule{\columnwidth}{0.4pt}

\twocolumngrid

\newpage

\begin{figure}[tb]
\centering
\includegraphics[width=\figurewidth]{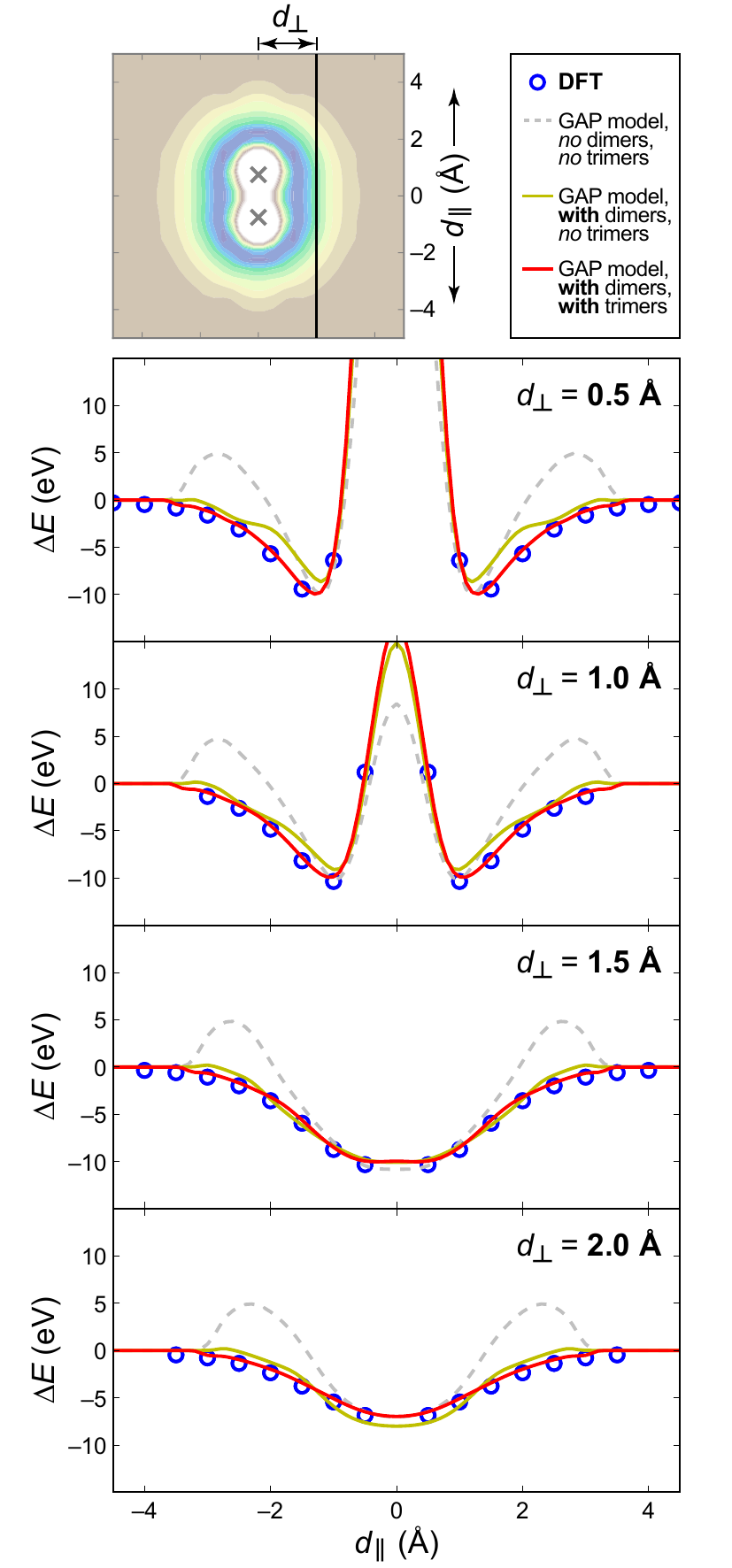}
\caption{\label{fig:SI_trimer_slices}
Line cuts through the datasets shown in the above colormaps: relative
energies are given for a line parallel to the dimer ($\times - \times$)
bond axis, and for different spacings perpendicular to this bond axis (see schematic
at the top left for definitions). The C--C distance in the central
dimer is 1.5 \AA{}.
Blue symbols show DFT data points, whereas lines denote GAP results;
these have been fitted using DFT data for the extended structures only
({\em dashed gray line}),
using a dataset including dimers but not trimers (as in the main text;
{\em yellow}), and finally using
a dataset that includes DFT data for both dimers and trimers ({\em red}).
}
\end{figure}

\newpage
\newpage

\end{document}